\documentclass[aps,prd,showpacs,twocolumn,superscriptaddress,nofootinbib,preprintnumbers]{revtex4-1} 

\makeatletter
\def\p@subsection{}
\makeatother

\usepackage{color,graphicx,amsmath,amssymb,lipsum}
\usepackage[colorlinks=true,citecolor=blue,linkcolor=blue,urlcolor=blue, backref=false,pdfborder={0 0 0}]{hyperref}

\usepackage[normalem]{ulem}
\usepackage[utf8]{inputenc} 
\usepackage{mathtools}

\usepackage{float}

\usepackage{ulem}
\usepackage{diagbox}

\usepackage[dvipsnames]{xcolor}

\newcommand{\be}{\begin{equation}}
\newcommand{\ee}{\end{equation}}
\newcommand{\beqa}{\begin{eqnarray}}
\newcommand{\eeqa}{\end{eqnarray}}

\newcommand\hMpc{h\text{Mpc}^{-1}}

\newcommand{\bseq}{\begin{subequations}}
\newcommand{\eseq}{\end{subequations}}

\renewcommand{\ln}{\mathop{\rm ln}\nolimits}

\newcommand\kmax{k_{\rm max}}

\def\gsim{\raise0.3ex\hbox{$\;>$\kern-0.75em\raise-1.1ex\hbox{$\sim\;$}}}
\def\lsim{\raise0.3ex\hbox{$\;<$\kern-0.75em\raise-1.1ex\hbox{$\sim\;$}}}

\def\beqn#1{\begin{equation}\label{#1}}
\def\eeqn{\end{equation}}

\def\beqa#1{\begin{eqnarray}\label{#1}}
\def\eeqa{\end{eqnarray}}

\def\Z2{$\mathcal{Z_2}$}

\newcommand {\ignore}[1]{}

\begin{document}

\preprint{INR-TH-2020-041}

\title{Constraints on the curvature of the Universe \\
and dynamical dark energy from the full-shape and BAO data}

\author{Anton Chudaykin}\email{chudy@ms2.inr.ac.ru}\affiliation{Institute for Nuclear Research of the
Russian Academy of Sciences, \\ 
60th October Anniversary Prospect, 7a, 117312
Moscow, Russia
}
\author{Konstantin Dolgikh}\email{dolgikh.ka15@physics.msu.ru}\affiliation{M.V. Lomonosov Moscow State University,\\
Vorobjevy Gory, 119991 Moscow, Russia
}
\author{Mikhail M. Ivanov}\email{mi1271@nyu.edu}\affiliation{Center for Cosmology and Particle Physics, Department of Physics, New York University, New York, NY 10003, USA}
\affiliation{Institute for Nuclear Research of the
Russian Academy of Sciences, \\ 
60th October Anniversary Prospect, 7a, 117312
Moscow, Russia
}

\begin{abstract} 
We present limits on the parameters of the o$\Lambda$CDM, $w_0$CDM, and $w_0 w_a$CDM models obtained from  
the joint analysis of the full-shape, baryon acoustic oscillations (BAO), big bang nucleosynthesis (BBN)
and supernovae data.
Our limits are fully independent of the data 
on the
cosmic microwave background 
(CMB) 
anisotropies, 
but rival the CMB constraints 
in terms of parameter error bars.
We find the spatial curvature consistent with a flat universe $\Omega_k=-0.043_{-0.036}^{+0.036}$ ($68\%$ C.L.);
the dark-energy equation of state parameter $w_0$ is measured to be $w_0=-1.031_{-0.048}^{+0.052}$ ($68\%$ C.L.), consistent with a cosmological constant.
This conclusion also holds for the time-varying dark energy equation of state,
for which we find
$w_0=-0.98_{-0.11}^{+0.099}$ and 
$w_a=-0.33_{-0.48}^{+0.63}$ (both at $68\%$ C.L.).
The exclusion of the supernovae data from the analysis does not significantly weaken our bounds.
This shows that using a single external BBN prior, 
the full-shape and BAO data can provide strong CMB-independent 
constraints on the 
non-minimal cosmological models.
\end{abstract}

\maketitle

\section{Introduction and Summary}

Elucidating 
the geometry of the Universe and  
the nature of the late-time expansion 
are some of the key goals of current and planned 
cosmological observations. 
The current data is consistent with the picture that the universe is flat 
and the late-time expansion can be described by a small cosmological constant~\cite{Aghanim:2018eyx}. 
These are the key assumptions of the base flat $\Lambda$CDM model,
whose parameters are accurately measured by the Planck CMB data~\cite{Aghanim:2019ame}.
The deviations from this model are strongly constrained
by the combination of the Planck data with baryon acoustic oscillations (BAO) and, optionally,
the supernovae data~\cite{Aghanim:2018eyx}. 

It has been recently shown that the parameters of the base $\Lambda$CDM model can 
be independently determined with the galaxy full-shape (FS) data\footnote{Our notion of the full-shape analysis 
should not be confused with the terminology of Ref.~\cite{Beutler:2016arn}, 
which studies how a fixed shape template gets distorted by the Alcock-Paczinsky effect~\cite{Alcock:1979mp}. 
In contrast, we use the power spectrum shape itself to constrain the physical cosmological parameters.}~\cite{Ivanov:2019pdj,DAmico:2019fhj,Colas:2019ret} collected by the Baryon Acoustic Oscillation Spectroscopic Survey (BOSS)~\cite{Alam:2016hwk}. 
This has become possible due to a
significant progress in large-scale structure theory achieved in the last decade with the 
development of the effective field theory of large scale structure (see ~\cite{Ivanov:2019pdj,DAmico:2019fhj} and references therein).
The full-shape data also sharpens the constraints on various extensions 
of the $\Lambda$CDM model: $\nu\Lambda$CDM, $\nu\Lambda$CDM+$N_{\rm eff}$~\cite{Ivanov:2019hqk,Philcox:2020xbv},
$w_0$CDM~\cite{DAmico:2020kxu}, and the early dark energy~\cite{Ivanov:2020ril,DAmico:2020ods}.
But crucially, it can even replace the CMB data in constraining beyond-$\Lambda$CDM scenarios.
An example is the minimal dynamical dark energy model $w_0$CDM~\cite{DAmico:2020kxu},
whose parameters can be determined from the big bang nucleosynthesis (BBN), BOSS FS, BAO and supernovae (SNe) data.
In this paper we continue testing non-minimal cosmological models with this data set, 
focusing on o$\Lambda$CDM, $w_0$CDM and $w_0w_a$CDM models. 

The main technical novelty of our analysis is the inclusion of the hexadecapole ($\ell=4$) moment of the redshift-space power spectrum, may break 
certain parameter degeneracies and yield stronger constraints on cosmological parameters. This is motivated by the result of the previous BOSS
full-shape
analysis from Ref.~\cite{Beutler:2016arn},
which have found that the hexadecapole 
yields a $\sim30\%$ improvement on the distance and RSD measurements. 
However, this result was obtained within the so-called alpha-parametrization, which does not assume any physical model for the late-time expansion. Hence, it is not 
clear if this improvement
will hold in
particular
physical models. 
For instance,
the posterior distribution
of the distance parameters 
obtained in the alpha-analysis
is significantly wider than 
the posterior space obtained 
in the $\Lambda$CDM 
model
\cite{Ivanov:2019hqk}.
However, 
the models that we consider here are characterized 
by several extra parameters controlling 
the late-time expansion, and hence their extended parameter space may be large enough 
to match the posterior 
distribution
sampled in the alpha-analysis.

Surprisingly, we found that 
this does not happen, i.e. 
the inclusion of the hexadecapole 
moment does not appreciably
narrow
the constraints 
on the extended models which
we consider here.
This suggests that improvement reported in Ref.~\cite{Beutler:2016arn} 
may be an artifact the alpha-analysis, probing the regions of parameter space which is unphysical
in the context of considered 
models.
A similar picture 
was found 
earlier in Ref.~\cite{Ivanov:2019hqk} in the context of the
$\Lambda$CDM model.



Deriving CMB-independent constants on the o$\Lambda$CDM, $w_0$CDM and $w_0w_a$CDM models 
is important for multiple reasons. 
The CMB data already provided tight constraints on the parameters of these models.
However, the CMB temperature likelihoods 
are known to be affected by various anomalies. 
In particular, 
the large-scale part of the spectrum 
exhibits the so-called ``low-$\ell$ deficit'' - suppression of the power for angular multipole numbers $20 \lesssim \ell\lesssim 30$. 
Besides that, 
late-time matter clustering determines the lensing smoothing of the acoustic peaks, 
whose observed amplitude is known to exceed the prediction of the $\Lambda$CDM model by over $2\sigma$~\cite{Aghanim:2018eyx}.
This is the so-called ``lensing anomaly'', which prefers 
models with enhanced large-scale structure growth, e.g. a Universe
with a positive spatial curvature~\cite{Aghanim:2018eyx,DiValentino:2019qzk}.
These anomalies have been intensely investigated in the past works, which showed 
that most likely they are just statistical fluctuations~\cite{Ade:2015xua,Aghanim:2018eyx,Addison:2015wyg,Aghanim:2016sns}. 
Nevertheless, the presence of these anomalies makes it desirable to have 
additional constraints from independent 
data sets.



There are a number of works that place the CMB-independent constrains on the considered
cosmological models. 
For instance, the eBOSS collaboration has recently reported $w_0=-0.69\pm0.15$
and $\Omega_k=0.078^{+0.086}_{-0.099}$ from the BAO data alone
\cite{Alam:2020sor},
whilst the DES analysis of galaxy clustering and weak gravitational lensing yielded $w_0=-0.82^{+0.21}_{-0.20}$ \cite{Abbott:2017wau}. 
We will show that the full-shape
data is able to significantly
improve upon these (and other) CMB-independent bounds.

In this paper, we infer the parameters of the o$\Lambda$CDM, $w_0$CDM and $w_0w_a$CDM models 
from a joint fit to the BOSS DR12 full shape data, supplemented with 
the BBN prior on the physical baryon density $\omega_b$, 
the BAO data from BOSS and eBOSS, and the Pantheon type Ia supernovae (SNe) measurements.
Our main result is that this data set is able to strongly constrain
the parameters of the considered non-minimal models:
\be
\begin{split}
&~~~\Omega_k= -0.043_{-0.036}^{+0.036}\,,\quad \text{o$\Lambda$CDM, FS+BAO+SNe}\,,\\
&~~~w_0 =
-1.031_{-0.048}^{+0.052}\,,\quad w_0\text{CDM, FS+BAO+SNe}\,,\\
&\begin{dcases} 
 w_0=-0.98_{-0.11}^{+0.10} \\ 
w_a=-0.32_{-0.48}^{+0.63}           
\end{dcases}  \quad w_0w_a\text{CDM, FS+BAO+SNe}\,.
\end{split}
\ee
Our limit on the spatial curvature of the Universe is comparable 
to the Planck TT+lowE measurement, $\Omega_k=-0.056^{+0.028}_{-0.018}$.
However, it is significantly weaker than the combined Planck+BAO+SNe limit. Still, our constraint on $\Omega_k$ is one of the strongest CMB-independent
bounds present 
in the literature.

The FS+BAO+SNe data set is very competitive with the CMB for the dynamical dark energy models.
Our error bars on 
the parameters
$w_0$ and  
$w_0-w_a$   
are only $\sim 30\%$ weaker  
than those from the Planck CMB + BAO + SNe data analysis~\cite{Aghanim:2018eyx}:
$w_0 =-1.028 \pm 0.032$ ($w_0$CDM), 
$w_0 = -0.961\pm 0.077,~w_a = -0.28^{+0.31}_{-0.27}$ ($w_0w_a$CDM).

The key ingredient of our analysis
is the BOSS full-shape likelihood introduced in Ref.~\cite{Ivanov:2019pdj}. 
Given the BBN prior on $\omega_b$,
the shape of the galaxy power spectrum provides us with a geometry-independent
constraint on the physical dark matter density~$\omega_{cdm}$. 
This fixes the sound horizon at decoupling and allows us to extract the geometric 
distances from the BAO measurements. 
These distances can be converted into
the parameters controlling the expansion history: the Hubble constant $H_0$, 
the effective spatial curvature density fraction $\Omega_k$,
the dark energy 
abundance $\Omega_{\rm de}$, along with the equation of state parameters $w_0, w_a$.
We have found nearly the same value of the 
sound horizon at the drag epoch $r_d$ in all
models that we consider in this work:
\be 
r_d=(146\pm 2.4)~\text{Mpc}\,.
\ee 

Remarkably, placing strong constraints on the expansion history is possible 
even without the SNe data. 
In particular, the BAO+FS measurements from BOSS and eBOSS, supplemented with a single
BBN prior, are enough to define the parameters of the $w_0$CDM and $w_0w_a$CDM models,
\be
\begin{split}
&~~~w_0 = -1.038_{-0.082}^{+0.1}\,,\quad w_0\text{CDM, FS+BAO}\,,\\
&\begin{dcases} 
    w_0 =-0.81_{-0.34}^{+0.25}\\
    w_a=-0.94_{-0.83}^{+1.3},             
\end{dcases}  \,, \quad w_0w_a\text{CDM, FS+BAO}\,.
\end{split}
\ee

All in all, our parameters limits for the $w_0$CDM and $w_0w_a$CDM models are comparable to those 
obtained from the combination of the Planck CMB, BAO and SNe data, whilst the $\Omega_k$
constraint is competitive with the primary Planck result, but is weaker than the full Planck + BAO limit.
The limits presented in this paper 
are some of the strongest CMB-dependent constraints on the o$\Lambda$CDM,
$w_0$CDM,
$w_0 w_a$CDM models.

It is important to note that the full-shape data allows 
us to accurately determine all relevant parameters 
of the o$\Lambda$CDM, $w_0$CDM and $w_0w_a$CDM models. In particular, 
we find the present-day Hubble constant $H_0$
consistent with the Planck base $\Lambda$CDM value~$H_0\simeq 68~$km/s/Mpc, 
with few percent error bars.
The optimal value of $H_0$ is very robust
to the considered extensions of $\Lambda$CDM, 
which shows that it can be accurately
measured from the full-shape and the BAO data
in a nearly model-independent way.

The remainder of this paper is structured as follows. 
We start with the discussion of our data sets in Sec.~\ref{sec:data}.
Sec.~\ref{sec:res} contains our main results. 
Finally, we
draw conclusions in Sec.~\ref{sec:concl}.
We present the validation of our pipeline on mock 
catalogs in Appendix~\ref{app:mocks}. 
Details of the analysis including
the Planck CMB data 
are presented in App.~\ref{app:planck}.

\section{Data and Methodology}
\label{sec:data}

\subsection{Data sets}

\textbf{Full-shape.} We use the multipoles of the redshift-space power spectrum of the luminous
red galaxies observed by BOSS~\cite{Alam:2016hwk}. 
The power spectrum multipoles were measured from the publicly available catalogs with the \texttt{nbodykit} code~\cite{Hand:2017pqn}.
The full-shape data is split in four non-overlapping chunks: low-z and high-z, 
north and south galactic caps. The effective redshifts are $z_{\rm eff}=0.38$ for the low-z samples and 
$z_{\rm eff}=0.61$ for the high-z samples. We use the data cuts $[0.01,~0.2]~h/$Mpc, which are
robust w.r.t. higher-order nonlinear corrections omitted in our theory model, see Appendix~\ref{app:mocks} for more detail.
We fit the full-shape data using one-loop perturbation theory implemented in the \texttt{CLASS-PT} code~\cite{Chudaykin:2020aoj}. We use covariance matrices from Patchy mocks~\cite{Kitaura:2015uqa}, which were shown to be robust w.r.t. stochastic noise biases~\cite{Wadekar:2020rdu,Philcox:2020zyp}. Further details on our theory models, covariance matrices and the window function treatment can be found in Refs.~\cite{Ivanov:2019pdj,Chudaykin:2020aoj,Wadekar:2020rdu}. 
Compared to these works, we also include the hexadecapole moment in our analysis.
We present the details of our hexadecapole treatment and validation on mocks in Appendix~\ref{app:mocks}.

\textbf{BAO.} We use the BAO measurements from the post-reconstructed 
power spectra of the BOSS DR12 data~\cite{Beutler:2016ixs}, which are covariant with 
the full-shape data from the pre-reconstruction power spectrum. 
We analyze these data sets with the methodology of Ref.~\cite{Philcox:2020vvt}.
Namely, we compute the anisotropic BAO parameters 
from mock catalogs and the BOSS data using the theoretical error approach~\cite{Baldauf:2016sjb}.
Then auto-covariance of the BAO parameters and their cross-covariance with 
the power spectrum multipoles is estimated from Patchy mocks~\cite{Kitaura:2015uqa}.

Additionally, we use the small-z BAO measurements from 6DF ($z_{\rm eff}=0.106$)~\cite{Ross:2014qpa} and SDSS DR7 MGS ($z_{\rm eff}=0.15$)~\cite{Beutler:2011hx}, 
along with the high redshift ($z_{\rm eff}=2.33$) Lyman-$\alpha$ forest auto-correlation
and 
the cross-correlation with quasars from eBOSS DR16~\cite{duMasdesBourboux:2020pck,Alam:2020sor}.
For completeness, 
we also use the BAO measurements from the 
eBOSS quasar sample ($z_{\rm eff}=1.48$)~\cite{Neveux:2020voa} and the emission line galaxy sample ($z_{\rm eff}=0.845$)~\cite{deMattia:2020fkb},
even though their impact on the eventual parameter constraints is quite marginal.
We do not use the BAO from the eBOSS LRG sample~\cite{Gil-Marin:2020bct} because it overlaps with
the tail of the BOSS DR12 high-z galaxy sample, which is already 
contained in
our joint full-shape-BAO
likelihood for this data chunk.

\textbf{Supernovae.} We will use the cosmological supernovae~Ia 
measurements from the Pantheon sample~\cite{Scolnic:2017caz}.

\textbf{BBN.} We will use the
BBN measurements from helium and deuterium~\cite{Aver:2015iza,Cooke:2017cwo} 
that constrain the
current physical  
density of baryons $\omega_b$. Specifically, we use the results of the ``standard'' analysis 
(see footnote 14 of Ref.~\cite{Ivanov:2019pdj} for more detail), implemented as a
following Gaussian prior:
\be
\omega_b\sim \mathcal{N}(0.02268,0.00038^2)\,. 
\ee

\subsection{Models}

We will consider three extensions of the $\Lambda$CDM model: o$\Lambda$CDM, $w_0$CDM and $w_0w_a$CDM, 
in the notation of Refs.~\cite{Alam:2020sor}. These models share the following set of parameters:\footnote{Note that we fix the current CMB monopole temperature $T_0$ to the COBE/FIRAS best-fit value $T_0=2.7255$ K~ \cite{Fixsen:1996nj}. 
$T_0$ has to be specified because it is an input 
parameter in the Boltzmann code \texttt{CLASS} that we use here~\cite{Blas:2011rf}.  
This choice is not crucial for our analysis. 
$T_0$ is irrelevant for the thermal history, but affects the late-time expansion through the contribution to the Friedman equation~\cite{Ivanov:2020mfr}. Given that, in principle, we could measure $T_0$ from the full-shape
and BAO data, but we prefer to use the FIRAS prior because it is very robust and 
has been independently confirmed by other probes.
}
\be
\label{eq:parcom}
\{\omega_b,\omega_{cdm},h, A_s,n_s\}\,, 
\ee
where $\omega_b,\omega_{cdm}$ are the current 
physical densities of baryons and dark matter, 
$h$ is the dimensionless Hubble constant ($H_0=h\cdot 100$ \mbox{km/s/Mpc}),
whereas $A_s$ and $n_s$ are the amplitude and tilt  
of the power spectrum of primordial scalar fluctuations.
Following~\cite{Aghanim:2018eyx}, we approximate the neutrino sector with one single massive state 
of mass $m_\nu=0.06~$eV.

The main difference between the models we consider 
shows up in the late-time expansion. The Friedman equation for these models read
\be
\begin{split}
& H^2=H_0^2\left(\Omega_m(1+z)^3+\Omega_\Lambda+\Omega_k(1+z)^2\right)\quad (\text{o$\Lambda$CDM})\\
& H^2=H_0^2\left(\Omega_m(1+z)^3+\Omega_{\rm de}(1+z)^{3(1+w_0)}\right)\quad (\text{$w_0\Lambda$CDM})\\
& H^2=H_0^2\left(\Omega_m(1+z)^3+\Omega_{\rm de}(1+z)^{3\left(1+w_0+\frac{w_a z}{(1+z)}\right)}\right)\\
&\quad \quad \quad \quad ~~~~~~~~~~~~~~~~~~~~~~~~~~~~~~~~~~~~~~~~~~~(\text{$w_0w_a\Lambda$CDM})
\end{split}
\ee
where $\Omega_m, \Omega_\Lambda, \Omega_{\rm de}$ are the current energy fractions of matter, cosmological constant
and dynamical dark energy, respectively, whereas $\Omega_k$ is the effective energy fraction of the spatial 
curvature. $\Omega_k>0$ corresponds to an open universe (negative curvature),
$\Omega_k<0$ describes a closed universe (positive curvature).
We will jointly fit the combined likelihood with all 
relevant cosmological parameters~\eqref{eq:parcom}, supplemented with 
$\Omega_k$ (for o$\Lambda$CDM), $w_0$ (for $w_0$CDM) and $w_0,w_a$ ($w_0w_a$CDM).

\subsection{Role of the full-shape data}

Before presenting results of our analysis, let us briefly discuss the role of the full-shape data.
To that end, it is instructive to start with the example of the minimal $\Lambda$CDM model. 
The shape of the galaxy power spectrum yields measurements 
of $\omega_b$ and $\omega_{cdm}$ through the scale-independent features, such as the relative 
ratios of the BAO peaks to the broadband slope~\cite{Mukhanov:2003xr,Tegmark:2006az,Ivanov:2019pdj}.
Even though a measurement of $\omega_b$ from the full-shape data is, in principle, possible,
the current limits are quite loose~\cite{Philcox:2020xbv,Ivanov:2019hqk}, which is why we prefer to use the BBN prior on 
$\omega_{b}$. 
Combining it with the shape, one can measure the comoving sound horizon at the drag epoch, which can be expressed as~\cite{Aubourg:2014yra},
\be
r_d=\frac{
55.148
}{(\omega_{cdm}+\omega_b)^{0.25351}\omega_b^{0.12807}}~\text{Mpc}\,. 
\ee
The sound horizon helps break the degeneracy with the distance to the galaxies present
in the angular scale of the acoustic horizon $\theta_{\rm BAO}=r_d/D_V(z_{\rm gal})$ and the angle 
of the matter-radiation equality scale \mbox{$\theta_{\rm eq}=1/(k_{\rm eq}D_V(z_{\rm gal}))$}, where 
$k_{\rm eq}$ is the conformal momentum of the perturbations entering the horizon at the matter-radiation equality, 
and
$D_V$
is the so-called comoving volume-averaged distance~\cite{Alam:2016hwk,Alam:2020sor}.
In the $\Lambda$CDM model the only remaining undetermined parameter is the Hubble constant,
and hence the measurement of $D_V$  
places a bound directly 
on $H_0$. 

The situation becomes more complicated when we consider extensions of $\Lambda$CDM
with more parameters defining the late-time background expansion. In this case, there appears
a geometric degeneracy which is similar to the geometric degeneracy of the CMB spectrum~\cite{Zaldarriaga:1997ch}. In principle, this degeneracy can be broken at the level of the full-shape data 
by means of the Alcock-Paczynski 
effect~\cite{Alcock:1979mp}, but this is a rather weak effect~\cite{Ivanov:2019pdj}. 
In particular it can only help constrain
the background parameters of the $\Lambda$CDM model with BAO 
when multiple redshifts are combined~\cite{Aubourg:2014yra,Cuceu:2019for,Schoneberg:2019wmt,Aghanim:2018eyx}. 
However, when we combine the full-shape with the BAO data at different redshifts (assuming a BBN prior), 
the acoustic horizon is fixed by the shape such that 
all the distance information
can be translated into the remaining background parameters. In the context of the $w_0$CDM model this has been
explicitly demonstrated in Ref.~\cite{DAmico:2020kxu}.

In principle, some information on the expansion history also comes form the large-scale structure 
growth, but its measurements are quite uncertain due to the galaxy bias and large cosmic
variance errors on the quadrupole and hexadecapole measurements~\cite{DAmico:2020kxu}.
Therefore, the bulk of our constraints is coming from the power spectrum monopole, which gives us $\omega_{cdm}$
and $D_V$.
Hence, we expect our limits to be robust w.r.t. possible contamination by the 
tensorial anisotropic assembly bias~\cite{Obuljen:2020ypy}, which 
affects the quadrupole and hexadecapole data. We leave a more detailed analysis of this effect for future work.

We stress that the crucial piece of information is the full-shape measurement
of $\omega_{cdm}$, which does not depend on the expansion model. 
However, it is affected by other parameters defining
the shape, e.g. the primordial power spectrum tilt $n_s$.
Given this reason, we vary this parameter in our analysis, even though the 
eventual constraints are $\sim$ x10 weaker than those from the Planck CMB data.
Still, we believe that it is more appropriate to treat $n_s$ as a free parameter
for the purpose of our analysis to derive the constraints independent from the CMB anisotropies.


\section{Results}
\label{sec:res}

In this section we present the results of our MCMC analyses 
for the considered models. In what follows we will suppress 
BBN in the notations 
of our data set, assuming that 
it is always included by default, i.e.
FS+BAO should be understood
as FS+BAO+BBN. In Sec. \ref{sec:planck} we supplement our baseline analysis with the full Planck likelihood.

\begin{table*}[t!]
  \begin{tabular}{|c||c|c||c|c||c|c|} \hline
   \diagbox{ {\small Param.}}{\small Dataset}  
   &  FB, o$\Lambda$CDM 
   & {\small FBS, o$\Lambda$CDM }  
   & FB, $w_0$CDM   
   &  FBS, $w_0$CDM   
   & FB, $w_0w_a$CDM  
   & FBS, $w_0w_a$CDM 
      \\ [0.2cm]
\hline
$\omega_{cdm}$   
& $0.1269_{-0.011}^{+0.0087}$ 
& $0.1273_{-0.011}^{+0.0086}$
& $0.1236_{-0.0099}^{+0.0082}$
& $0.1233_{-0.0096}^{+0.0083}$
& $0.126_{-0.01}^{+0.0085}$
& $0.1252_{-0.01}^{+0.0085}$
\\ 
\hline 
  $h$   & $0.6948_{-0.016}^{+0.015}$
  & $0.6945_{-0.015}^{+0.013}$
  & $0.69_{-0.021}^{+0.018}$
  & $0.6885_{-0.014}^{+0.013}$
  &  $0.6775_{-0.026}^{+0.026}$
  & $0.6896_{-0.014}^{+0.013}$
  \\ \hline
$\ln(10^{10}A_s)$   
& $2.64_{-0.20}^{+0.20}$ 
& $2.65_{-0.20}^{+0.20}$ 
& $2.77_{-0.17}^{+0.17}$
& $2.77_{-0.16}^{+0.16}$
& $2.72_{-0.17}^{+0.17}$
& $2.72_{-0.17}^{+0.17}$
\\ 
\hline
$n_s$  
& $0.9153_{-0.067}^{+0.068}$
& $0.9136_{-0.065}^{+0.068}$
& $0.9334_{-0.064}^{+0.066}$
& $0.9346_{-0.064}^{+0.065}$
& $0.9191_{-0.065}^{+0.066}$
& $0.9236_{-0.065}^{+0.066}$
\\
\hline
$\Omega_k$   
& $-0.044_{-0.044}^{+0.043}$
& $-0.043_{-0.036}^{+0.036}$
& $-$
& $-$ 
& $-$
& $-$
\\   \hline 
$w_0$   
& $-$ 
& $-$
& $-1.038_{-0.082}^{+0.1}$
& $-1.031_{-0.048}^{+0.052}$
&
$-0.805_{-0.34}^{+0.25}$
&$-0.9826_{-0.11}^{+0.099}$ 
\\   \hline
$w_a$   
& $-$ 
& $-$
& $-$
& $-$ 
& $-0.9451_{-0.83}^{+1.3}$
&$-0.3264_{-0.479}^{+0.629}$
\\   \hline
\hline
$\Omega_\Lambda$   
& $0.733_{-0.044}^{+0.044}$
& $0.731_{-0.033}^{+0.033}$
& $-$
& $-$ 
& $-$ 
& $-$ 
\\ \hline
$\Omega_m$   
& $0.3109_{-0.014}^{+0.013}$
&  $0.3119_{-0.013}^{+0.012}$
& $0.3087_{-0.016}^{+0.015}$
& $0.3091_{-0.012}^{+0.011}$
& $0.3262_{-0.031}^{+0.023}$
& $0.3121_{-0.013}^{+0.012}$
\\ \hline
$\Omega_{\rm de}$   & $-$
& $-$
& $0.692_{-0.015}^{+0.015}$
& $0.691_{-0.011}^{+0.011}$
& $0.674_{-0.022}^{+0.031}$
& $0.688_{-0.012}^{+0.012}$
\\ \hline
$\sigma_8$   & $0.708_{-0.048}^{+0.043}$
&  $0.708_{-0.048}^{+0.043}$
& $0.718_{-0.048}^{+0.043}$ 
& $0.718_{-0.048}^{+0.043}$ 
& $0.705_{-0.049}^{+0.044}$
& $0.711_{-0.049}^{+0.044}$
\\ 
\hline
$r_d$ [Mpc]  & $146_{-2.4}^{+2.4}$
& $146_{-2.4}^{+2.4}$
& $146_{-2.4}^{+2.4}$ 
& $146_{-2.4}^{+2.4}$ 
& $146_{-2.4}^{+2.4}$
& $146_{-2.4}^{+2.4}$
\\ 
\hline
\end{tabular}
\caption{Mean values and 68\% CL minimum credible
intervals for the parameters of the various extended models for two data sets differing by the presence of the supernovae data. FB denotes the combination FS+BAO, FBS denotes the combination FS+BAO+SNe. The BBN prior on $\omega_b$ is assumed in all analyses, and the corresponding posterior is not displayed because it is prior-dominated.
The top group represents the parameters that were directly varied in the MCMC chains. The bottom group
are the derived parameters.}
\label{table1}
\end{table*}

\begin{figure*}[ht]
\begin{center}
\includegraphics[width=1\textwidth]{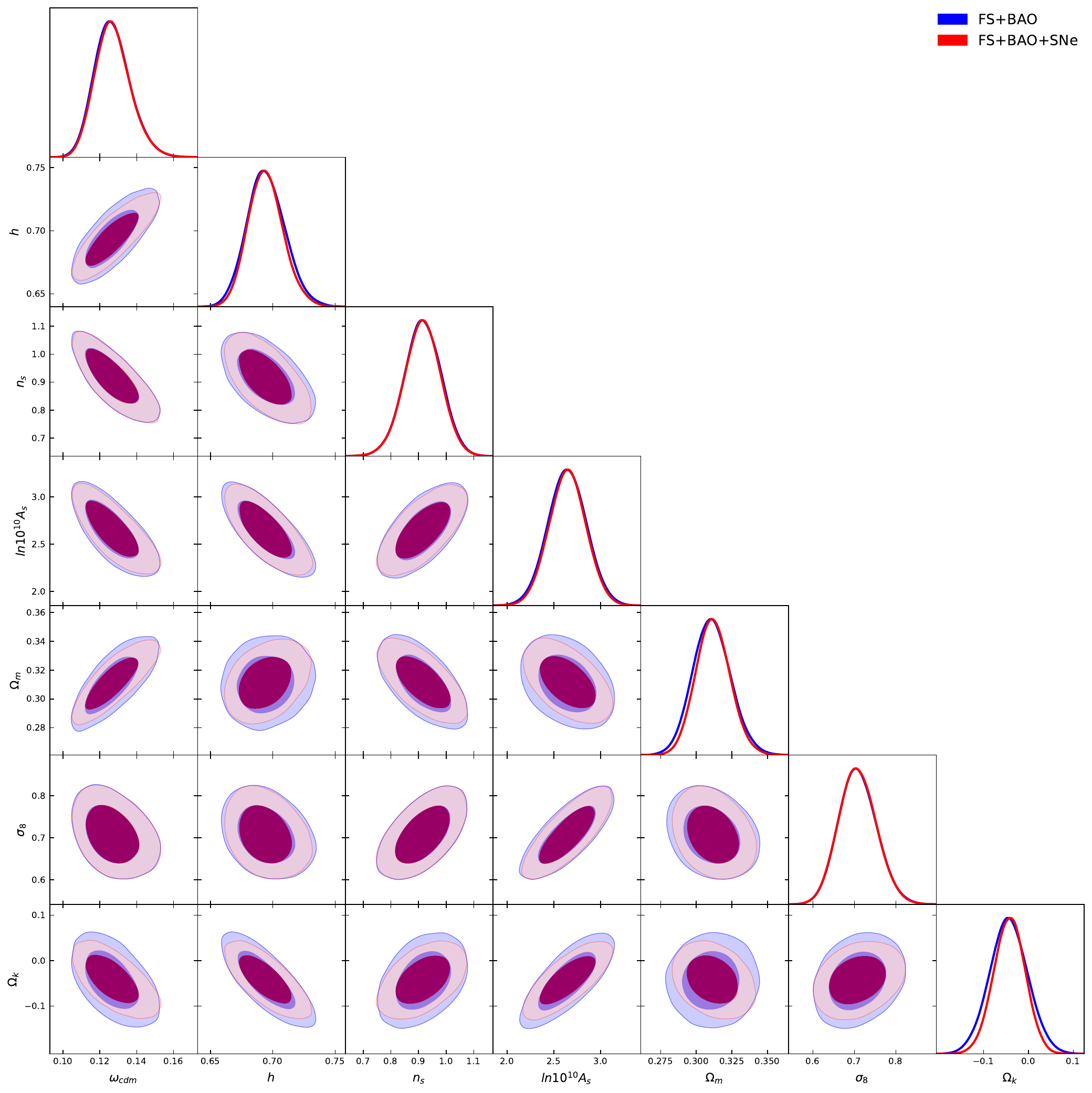}
\end{center}
\caption{
Posterior distributions of the cosmological parameters of the o$\Lambda$CDM model.
\label{fig:Ok} } 
\end{figure*}
\begin{figure*}[ht]
\begin{center}
\includegraphics[width=1\textwidth]{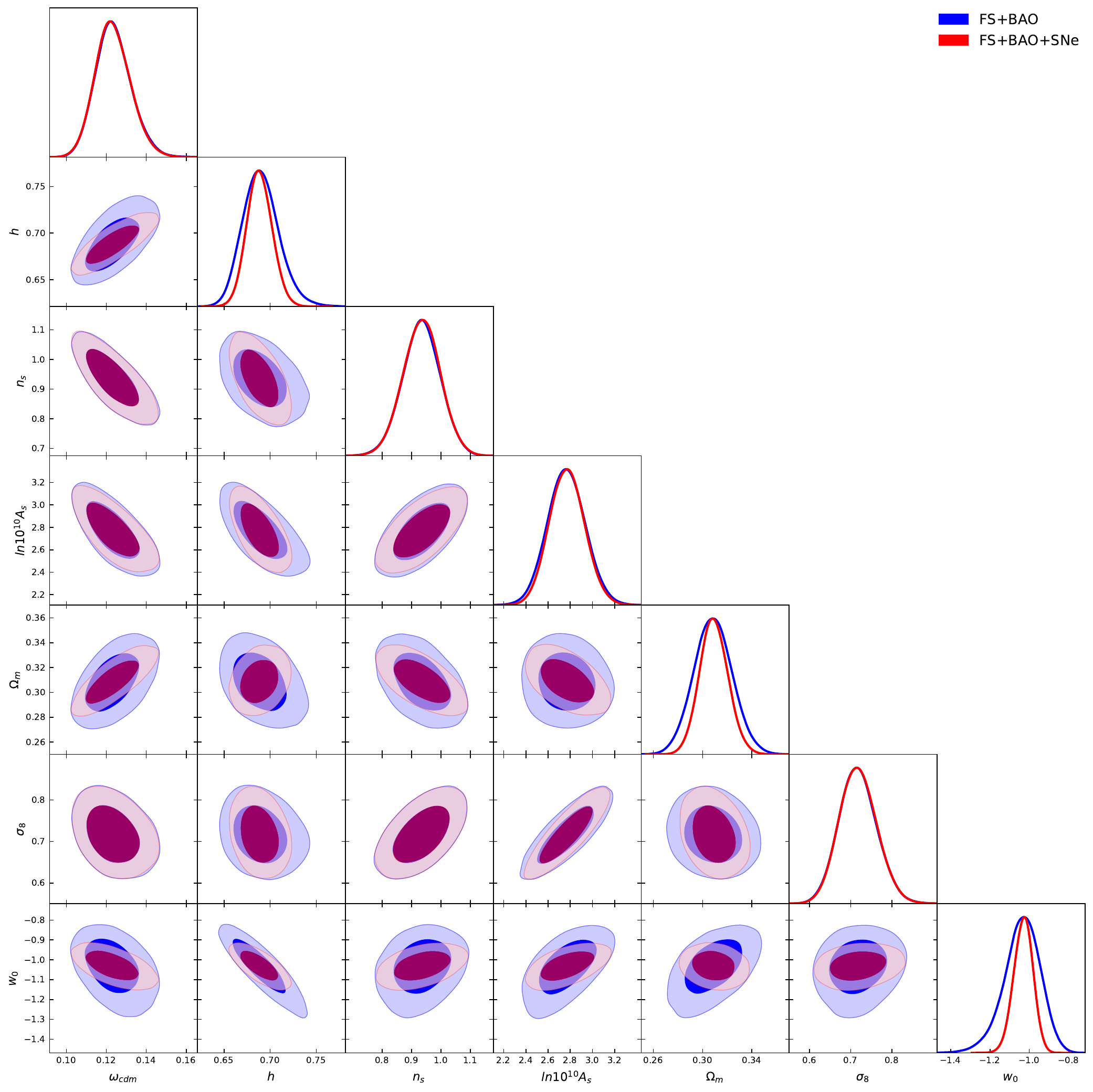}
\end{center}
\caption{
Posterior distributions of the cosmological parameters of the $w_0$CDM model.
\label{fig:w0} } 
\end{figure*}
\begin{figure*}[ht]
\begin{center}
\includegraphics[width=1\textwidth]{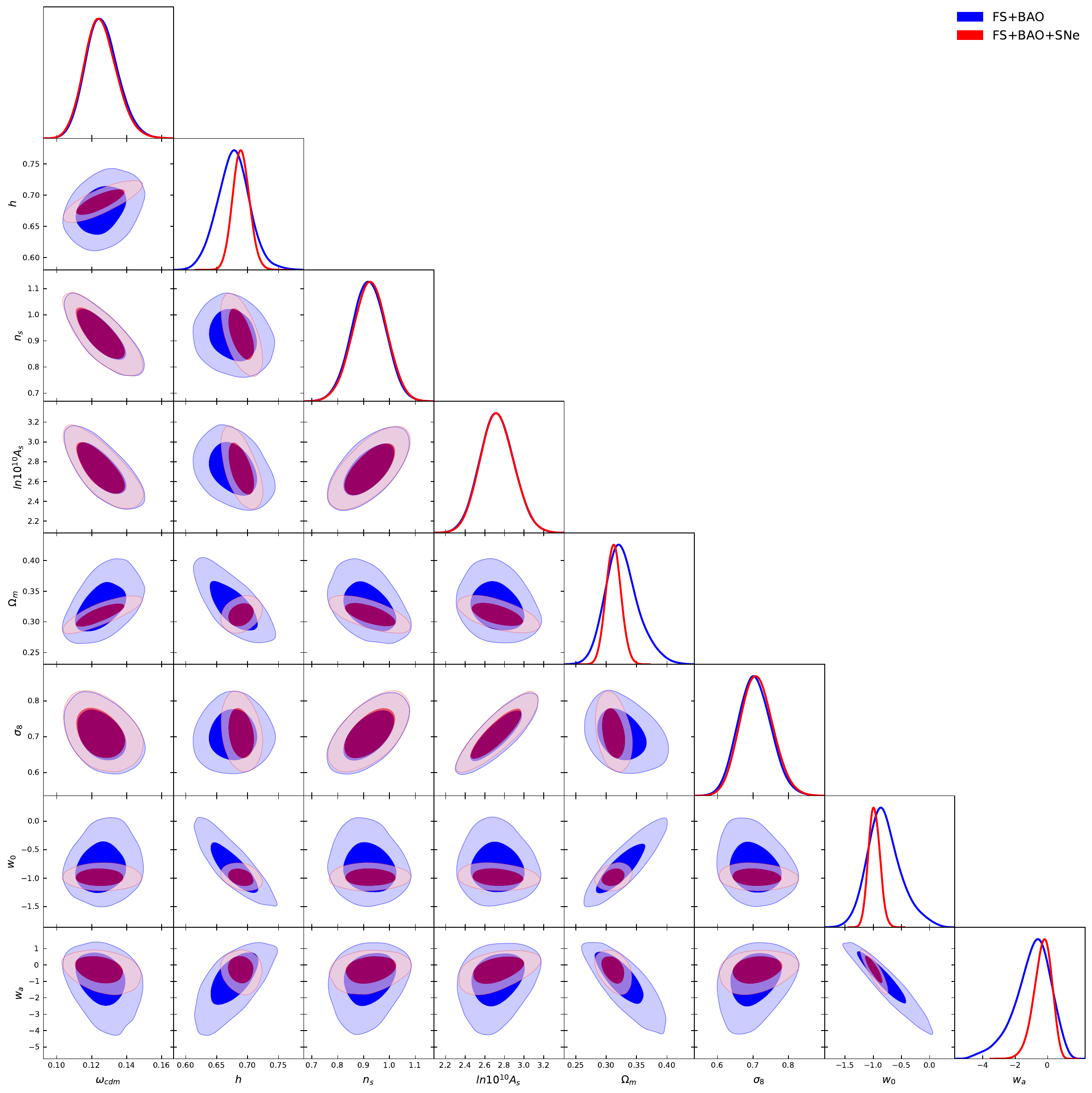}
\end{center}
\caption{
Posterior distributions of the cosmological parameters of the $w_0w_a$CDM model.
\label{fig:w0wa} } 
\end{figure*}

\subsection{o$\Lambda$CDM}

The triangle plot for the parameters of the o$\Lambda$CDM model is shown in Fig.~\ref{fig:Ok}, the 1d marginalized
constraints are presented in the 2nd and 3rd columns of Table~\ref{table1} for the FS+BAO and FS+BAO+SNe 
datasets. The first relevant observation is that the constraints on the shape and amplitude parameters $\omega_{cdm},n_s,A_s$ and $\sigma_8$
in the o$\Lambda$CDM model are similar to those from the flat $\Lambda$CDM model~\cite{Ivanov:2019pdj} (see also Appendix~\ref{app:mocks}).
This is consistent with the fact that the shape constraints do not depend on projection effects~\cite{Ivanov:2019hqk}. 
The second relevant observation is that the spatial curvature is consistent 
with zero within $95\%$CL. Remarkably, the FS+BAO and FS+BAO+SNe data 
yield a very significant evidence for the cosmological constant even in the presence 
of the non-zero spatial curvature in the fit.

Comparing the FS+BAO (`FB') and FS+BAO+SNe (`FBS') cases, we see that the addition of the SNe data improves 
the $\Omega_k$ constraint only by $\sim 20\%$. This shows that the curvature constraints
are indeed dominated by the FS+BAO data sets. 
Comparing our results to the BAO-only measurements of $\Omega_k$ and $\Omega_m$~\cite{Alam:2020sor}, we see that the addition of the FS data shrinks the error bars by a factor of $\sim 3$. Moreover, it allows us to measure $H_0$ to $2\%$ accuracy, 
which would not be possible with the BAO-only data.

Let us discuss the implications for the Planck spatial curvature measurements.
It is known that the primary Planck data favors the model with positive
spatial curvature; i.e. the Planck 2018 TT+low E likelihood prefers a closed 
universe with more than $2\sigma$ significance, $\Omega_k= -0.056^{+0.028}_{-0.018}$ \cite{Aghanim:2018eyx}.
Our measurement is consistent with this estimate, 
although the error bars are large enough to enclose the $\Omega_k=0$ within $2\sigma$ as well.
The Planck preference for positive spatial curvature can be traced back to the 
lensing anomaly, which is, most likely, just a statistical fluke~\cite{Aghanim:2018eyx}. Indeed, the combined
Planck+BAO+SNe dataset gives $\Omega_k= -0.0001\pm 0.0018$, consistent with the flat $\Lambda$CDM model.

It is worth mentioning that there exist other independent estimates of the spatial
curvature of the Universe from different combinations of 
the SH0ES distance ladder measurements~\cite{DiValentino:2019qzk},
strong lensing time-delays, BAO, BBN, 
cosmic chronometers and
quasar luminosity distances, see~\cite{DiValentino:2020srs} and references therein.
The strongest one is obtained
from the combination
BAO+SNe+BBN+SH0ES, $\Omega_k=-0.091\pm 0.037$~\cite{DiValentino:2019qzk}.
This limit is, however, crucially depends 
on the inclusion of the SH0ES $H_0$ prior~\cite{Riess:2019cxk}, which
is in tension with other data sets~\cite{DiValentino:2020zio}, and hence,
this limit should be taken with a grain of salt 
until the tension is resolved. 
The error bars from other
CMB-independent measurements
of $\Omega_k$
are at least four times larger than the uncertainty 
of our measurement.

\subsection{$w_0$CDM}

Now let us focus on the $w_0$CDM model. 
Planck-alone prefers very negative $w_0$.
The triangle plot for the parameters of the o$\Lambda$CDM model is shown in Fig.~\ref{fig:w0}, the 1d marginalized
constraints are presented in the 4th and 5th 
columns of Table~\ref{table1} for the FS+BAO and FS+BAO+SNe data.
The FS+BAO data yields the dark energy equation of state parameter
compatible with the cosmological constant value at $68\%$CL. This can be contrasted 
with the Planck TT+low E constraints (see chapter 17.1 of~\cite{legacy}) $w_0=-1.56^{+0.19}_{-0.39}$
preferring a $\sim 2\sigma$ shift of $w_0$ into the phantom domain. 

Our FS+BAO constraint is
almost twice stronger than the BAO-only result $w_0=-0.69\pm 0.15$~\cite{Alam:2020sor}, which is also shifted away from the cosmological constant prediction.

Our final constraints from the  FS+BAO+SNe data set
are somewhat weaker (by $\sim 30\%$), but still competitive with 
the Planck+BAO+SNe result $w_0=-1.028\pm 0.032$.
It is also useful to compare our results with the pioneering
analysis of the $w_0$CDM model~\cite{DAmico:2020kxu} with the FS+BAO+SNe data.
This analysis was based on 
the same data set as ours, the only difference is the 
addition of the hexadecapole moment in the present work, along with a small update
in the eBOSS BAO likelihood.
We observe that this reduces the error bar on $w_0$ by $\simeq 30\%$ 
in the FS+BAO case and by $\sim 10\%$ in the FS+BAO+SNe case.

\subsection{$w_0w_a$CDM}

Finally, let us discuss the $w_0w_a$CDM model. 
The triangle plot for the parameters of the o$\Lambda$CDM model is shown in Fig.~\ref{fig:w0wa}, the 1d marginalized
constraints are presented in the 6th and 7th 
columns of Table~\ref{table1} for the FS+BAO and FS+BAO+SNe data.

Our constraints on $w_0w_a$CDM model from FS+BAO are comparable to the constraints from
different combinations 
of the primary Planck data with the BAO~\cite{legacy}. 
In particular, the strongest combination including CMB lensing yields $w_0=-0.59\pm 0.27$, $w_a = -1.24\pm 0.74$, which is only $\sim 10\%$
better than our FS+BAO estimate.
The situation somewhat changes with the inclusion of the SNe, which noticeably shrinks
the constraints from Planck+BAO to the level $w_a= -0.961 \pm 0.077$, 
$w_a =  -0.28^{+0.31}_{-0.27}$. 
Our limits on $(w_0,w_a)$ from the FS+BAO+SNe data are 
$(20\%,40\%)$ weaker
than this result. 

\subsection{Combined analysis with Planck}
\label{sec:planck}

In this section we extend our baseline analysis by including the full Planck data. We compare cosmological parameters of the o$\Lambda$CDM, $w_0$CDM and $w_0w_a$CDM models from the FS+BAO+SNe+Planck data with that from FS+BAO+SNe alone. The aim of this analysis is to see to what extent the Planck CMB data can improve upon the FS+BAO+SNe constraints presented in the previous section. Since Planck provides
us with a measurement of 
the physical baryon density
that is better than that 
of BBN, we do not include 
the BBN prior in the Planck analysis.

We found that in the o$\Lambda$CDM model the 
constraints on all cosmological parameters are significantly improved after including the Planck data. Specifically, the error bar on $\Omega_k$ reduces by more than one order of magnitude. For the $w_0$CDM and $w_0w_a$CDM models we found  more modest improvements:
the Planck data 
only narrows the posteriors of 
shape 
and 
amplitude 
parameters ($\omega_{cdm},n_s,A_s$), whereas the constraints on the background 
expansion
parameters ($h$, $w_a$, $w_a$) are very similar to those from FS+BAO+SNe.
Further details of our analysis can be found in App.~\ref{app:planck}.


\section{Conclusions\label{sec:concl}}

In this paper we have presented the measurements of the parameters of o$\Lambda$CDM, $w_0$CDM and $w_0w_a$CDM models from the full-shape power spectrum data, supplemented with the BBN, BAO and SNe measurements. 
Our constraints on the parameters of these models are significantly better
than those based on the BAO data only, which clearly shows the statistical power of the effective field theory-based full-shape power spectrum likelihood~\cite{Ivanov:2019pdj,Chudaykin:2020aoj}. 
The measurements presented in this paper are also competitive with the Planck+BAO(+SNe)
limits, especially for the dynamical dark energy model. We also found that the Planck data only marginally improve the constraints on $w_0$ and $w_a$ parameters but significantly narrows the error bar on $\Omega_k$.

Importantly, 
the full-shape data 
allows us to place constraints on all relevant 
cosmological parameters of the considered non-minimal models. One of such parameters
is the present-day Hubble constant $H_0$, which we measure to (1-2)$\%$ precision 
even in the extensions of $\Lambda$CDM. 
Remarkably, our results 
agree with the Planck-preferred results. 
This is an important test, showing good agreement between
various data sets:
the CMB~\cite{Aghanim:2018eyx,Aiola:2020azj}, 
large-scale structure~\cite{Ivanov:2019pdj,DAmico:2019fhj,Schoneberg:2019wmt,Cuceu:2019for,Aubourg:2014yra,Abbott:2017smn}, 
the local 
measurements from the tip of the red giants branch~\cite{Freedman:2019jwv},
and strong lensing time-delays (after taking into account the mass-sheet degeneracy)~\cite{Birrer:2020tax}. 
These estimates, however, are still in tension with Cepheid-calibrated supernovae measurements, see~\cite{DiValentino:2020zio}
and references therein.

Our results have some implications for the so-called $\sigma_8-$tension~\cite{DiValentino:2020vvd},
the apparent 
disagreement on the value of $\sigma_8$ 
between Planck on one side and
various 
large scale structure measurements
on the other side, e.g.
weak lensing measurements by the Dark Energy Survey~\cite{Abbott:2017wau}
and Kilo-Degree Survey~\cite{2020arXiv200715632H}. 
In all models that we considered,
we found the mass fluctuation amplitude $\sigma_8$ systematically lower 
than the Planck predictions for the same models, 
although the significance of this tension
in terms of our error bars
is quite low ($< 2\sigma$).
In order to draw more robust conclusions we need to reduce the statistical error
of our measurement, which can be done either by including external data sets 
\footnote{e.g. the 
bispectrum~\cite{Gil-Marin:2014sta,Gil-Marin:2016wya}, the void-galaxy correlation~\cite{Nadathur:2020vld}, or the counts-in-cells statistic~\cite{Uhlemann:2015npz,Ivanov:2018lcg,Repp:2020kfd}.}, 
or collecting more data. The latter will certainly happen 
in the future with the Euclid~\cite{Laureijs:2011gra,Amendola:2016saw} and DESI~\cite{Aghamousa:2016zmz} surveys, which promise to dramatically sharpen the 
precision of cosmological parameter measurements, see e.g.~\cite{Chudaykin:2019ock,Audren:2012vy,Brinckmann:2018owf,Orsi:2009mj,Yankelevich:2018uaz}.

Overall, we have found no evidence for any of the extensions
of the base $\Lambda$CDM model in our analysis of the FS+BAO+SNe data, which is fully 
independent from the Planck CMB anisotropies. 
Our analysis confirms a remarkably concordant picture of the universe,
whose properties on a wide range of redshifts can be described within 
the simple flat $\Lambda$CDM model.

\vspace{1cm}
\section*{Acknowledgments}

We thank Oliver Philcox
and Marko Simonovic for valuable discussions.
The work is supported by the RFBR grant 20-02-00982. 
Our numerical calculations were partially performed with the HybriLIT heterogeneous computing platform (LIT, JINR) (\href{http://hlit.jinr.ru}{http://hlit.jinr.ru}).

Parameter estimates presented in this paper are obtained with the \texttt{CLASS-PT} 
Boltzmann code \cite{Chudaykin:2020aoj}
(also see \cite{Blas:2011rf}) interfaced with the \texttt{Montepython} MCMC sampler~\cite{Audren:2012wb,Brinckmann:2018cvx}. 
The plots with posterior densities and marginalized limits are generated with the latest version of the \texttt{getdist} package\footnote{\href{https://getdist.readthedocs.io/en/latest/}{
\textcolor{blue}{https://getdist.readthedocs.io/en/latest/}}
}~\cite{Lewis:2019xzd},
which is part of the \texttt{CosmoMC} code~\cite{Lewis:2002ah,Lewis:2013hha}. 

We are grateful to H\'ector Gil-Mar\'in 
for making the Nseries mocks and related
data products publicly available~\cite{Gil-Marin:2015sqa,hector}.

\appendix 

\section{Details of the full-shape likelihood and the hexadecapole moment}
\label{app:mocks}

\begin{table*}[t!]
  \begin{tabular}{|c||c|c|} \hline
   \diagbox{ {\small Param.}}{\small Dataset}  
   &  BOSS volume
   & 10xBOSS volume  
 \\ [0.2cm]
\hline
$\Delta\omega_{cdm}/\omega_{cdm}$   
& $0.0370_{-0.11}^{+0.086}$
& $0.0151_{-0.046}^{+0.037}$
\\ 
\hline 
  $\Delta h/h$   & $0.00599_{-0.017}^{+0.019}$
  &$0.00192_{-0.0068}^{+0.0069}$ 
  \\ \hline
$\Delta n_s/n_s$  
& $-0.015_{-0.074}^{+0.074}$
& $-0.0039_{-0.028}^{+0.031}$
\\
\hline
$\Delta A_s/A_s$   & $-0.0243_{-0.18}^{+0.12}$ 
& $0.0239_{-0.070}^{+0.070}$ 
\\ 
\hline\hline
$\Delta\ln(10^{10}A_s)/\ln(10^{10}A_s)$   & $-0.0119_{-0.053}^{+0.053}$ 
& $0.0069_{-0.022}^{+0.022}$ 
\\ 
\hline
$\Delta\Omega_m/\Omega_m$   & $0.0181_{-0.072}^{+0.060}$
& $0.0086_{-0.029}^{+0.024}$
\\ \hline
$\Delta \sigma_8/\sigma_8$   & $-0.0016_{-0.062}^{+0.056}$
&$0.0188_{-0.022}^{+0.022}$ 
\\ 
\hline
\end{tabular}
\caption{Mean values and 68\% CL minimum credible
intervals for the parameters of $\Lambda$CDM model inferred from the PT Challenge
simulation spectra at $z=0.61$
for two choices of the covariance matrix, corresponding to the cumulative BOSS volume (2nd column)
and 10 times the BOSS volume (3rd column).
We display all parameters as $(p-p_{\rm fid.})/p_{\rm fid.}$, where $p_{\rm fid.}$
is the fiducial value used 
in simulations.
The top group represents the parameters that were directly varied in the MCMC chains. The bottom group
are the derived parameters.}
\label{tab:PTC}
\end{table*}
\begin{figure*}[ht]
\begin{center}
\includegraphics[width=1\textwidth]{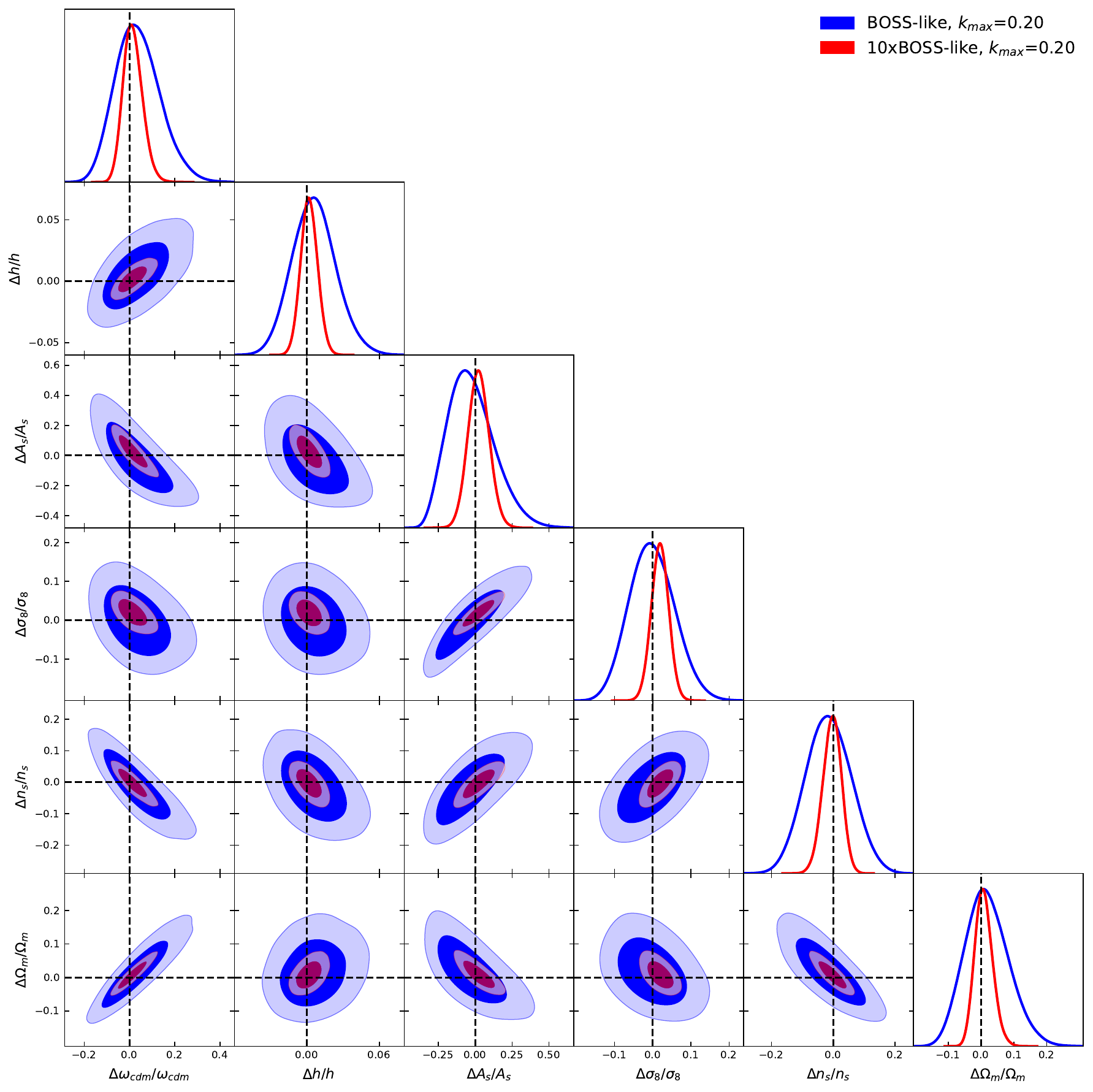}
\end{center}
\caption{
Posterior distributions of the cosmological parameters of the $\Lambda$CDM model fitted to the PT Challenge mock data~\cite{Nishimichi:2020tvu}. All parameters are normalized to represent the relative
deviations from the fiducial values used in the N-body simulations.
\label{fig:PTC} } 
\end{figure*}

In this Appendix we 
discuss in detail 
the effective-field 
theory based 
full-shape power 
spectrum likelihood.
We use the likelihood introduced in  Ref.~\cite{Ivanov:2019hqk}, but make several changes compared to the analysis of this paper. First, we use slightly different priors on the nuisance parameters. 
Second, we include 
the power spectrum hexadecapole moment $\ell=4$. 
Third, as a result of including the hexadecapole, the posterior parameter volume shrinks, and the theory-systematic 
error becomes more sizeable. 
This motivates us to use a more
conservative data cut compared to Ref.~\cite{Ivanov:2019hqk}.
In what follows we validate the
priors and data cuts
used in our analysis.

The rationale behind the inclusion
of the hexadecapole is the following one. The models that we consider
in this paper mainly 
alter the background
expansion and the growth of structures. 
These are probed through the distance measurements and redshift-space distortions~\cite{Alam:2016hwk}. 
The usual method to extract these quantities  
from the redshift-space power spectrum
is the so-called alpha analysis, see Ref.~\cite{Beutler:2016arn} and Ref.~\cite{Ivanov:2019hqk}
for a justification of this analysis in certain contexts.
Ref.~\cite{Beutler:2016arn} has shown that the inclusion
of the hexadecapole improves the distance and RSD measurements by $\sim 30\%$. Therefore, we expect 
that the hexadecapole should also shrink 
the parameter contours in the extended models that we consider here. 

The caveat, however, is that the alpha-analysis 
does not rely on any physical model of the late-time expansion (though an early-universe model is still required
in order to generate the power spectrum template). 
Therefore, the alpha-analysis does not respect 
relations between the radial and angular distances, which exist in particular models. 
This means that parts of the parameter space probed by the alpha-analysis can be unphysical. 
An example of this situation is the $\Lambda$CDM model,
which, in fact, 
corresponds to an extremely narrow region of the 
parameter space probed by the alpha-analysis~\cite{Ivanov:2019hqk}. 
It can be that
the physical priors 
on the distances diminish the information gain 
from the hexadecapole.
Therefore, the natural question is whether the hexadecople
improves parameter constraints in the complete
full-shape analysis done at the level of the physical models. 
Addressing this question
is one of the goals of this section.

\subsection{Nuisance parameters and priors}

We fit the full-shape data with the
one-loop perturbation theory model that is described 
by the following set of nuisance parameters (see~\cite{Chudaykin:2020aoj}
for details):
\be 
\{b_1,b_2,b_{\mathcal{G}_2},b_{\Gamma_3},c_0,c_2,\tilde{c},P_{\rm shot},a_2\}\,,
\ee 
where $b_1$ is the linear galaxy bias, $b_2$ is the local-in-density quadratic bias, $b_{\mathcal{G}_2}$
and $b_{\Gamma_3}$ 
are the quadratic and cubic
tidal biases, $c_0$ and $c_2$ are the higher derivative biases for the monopole 
and quadrupole ($k^2$-counterterms), $\tilde{c}$ is the higher-order $k^4$ redshift-space counterterm, 
$P_{\rm shot}$ is the residual constant shot noise 
contribution and $a_2$
is the scale-dependent 
redshift-space stochastic 
counterterm~\cite{Perko:2016puo},
which we define as
\be 
\label{eq:a2}
P_{\rm stoch, RSD}=a_2\left(\frac{k}{k_{\rm NL}}\right)^2\frac{1}{\bar n}\,,
\ee 
where $\bar n$ is the galaxy number-density 
and $k_{\rm NL}$
is the non-linear scale. 
The $a_2$ counterterm was not used in Ref.~\cite{Ivanov:2019hqk} because it was found to be fully degenerate with $\tilde{c}$ at the level of the monopole and quadrupole moments. The hexadecapole moment breaks this degeneracy. Even though we do not detect this coefficient, we prefer to scan over it in our MCMC chains because it affects the parameter error bars. We use the following priors on the nuisance parameters: 
\be 
\begin{split}
& b_1A^{1/2}\in \text{flat}[1,4],\quad
b_2A^{1/2}\sim \mathcal{N}(0,1^2),\\
& 
b_{\mathcal{G}_2}A^{1/2}\sim \mathcal{N}(0,1^2),
\quad b_{\Gamma_3}\sim \mathcal{N}(0.65,1^2), \\
& c_0\sim \mathcal{N}(0,30^2),\quad 
c_2\sim \mathcal{N}(30,30^2)\\
& \tilde{c}
\sim \mathcal{N}(500,500^2),\quad 
P_{\rm shot}
\sim 
\mathcal{N}(0,5\cdot 10^3)\,,
\end{split}
\ee 
where $A\equiv A_s/A_{s,\,{\rm fid.}}$ (see Eq.~\eqref{eq:A}).
The physical motivation behind the choice of our priors can be found in Refs.~\cite{Chudaykin:2020aoj,Wadekar:2020rdu}.
As far as $a_2$ is concerned, we set the following physical prior, see Eq.~\eqref{eq:a2}:
\be 
\begin{split}
&a_2\sim \mathcal{N}(0,\,2^2)\,,
\text{with}
\quad 
k_{\rm NL}=0.45~h\text{Mpc}^{-1},
\\
& \bar{n}^{-1}=5\cdot 10^3~[h^{-1}\text{Mpc}]^3\,.
\end{split}
\ee 
We set the scale-dependent stochastic counterterm \mbox{$a_0=0$} as suggested 
by the field level 
analysis of the BOSS-like dark matter halos~\cite{Schmittfull:2018yuk}.

Note that unlike the reference~\cite{Ivanov:2019hqk}, we marginalize over $b_{\Gamma_3}$ 
assuming a prior centered at the prediction of the coevolution 
model and with unit variance~\cite{Desjacques:2018pfv}.
Fixing $b_{\Gamma_3}$ or marginalizing over it does not have an impact on our constraints~\cite{Wadekar:2020rdu}. Nevertheless, we prefer to do a marginalization
over this unknown coefficient within a physically-motivated prior in order to be rigorous. 

\subsection{Validation on mock catalogs}

The pipeline used in our work 
was already validated in Refs.~\cite{Ivanov:2019hqk,Nishimichi:2020tvu}. However, these works did not include the hexadecapole moment, which 
can change the conclusions 
on the data cut $k_{\rm max}$
used in the analysis. To check this, we test our pipeline 
on mock catalogs of the BOSS-like luminous 
red galaxies in this section in two different regimes. As a first step, we will fit the mock 
data from the periodic box
N-body simulations 
PT Challenge (`perturbation theory challenge')~\cite{Nishimichi:2020tvu}.
As a second step, we will analyze the mock data from 
more realistic mock catalogs that include the survey
mask and selection functions. 

\begin{table*}[t!]
  \begin{tabular}{|c||c||c|c||c|c|} \hline
   Param.
   & fiducial
   &  $P_{0,2}~(k_{\rm max}=0.2)$
   & $P_{0,2,4}~(k_{\rm max}=0.2)$  &
   $P_{0,2}~(k_{\rm max}=0.25)$
   &$P_{0,2,4}~(k_{\rm max}=0.25)$
 \\ [0.2cm]
\hline
$\omega_{cdm}$   
&$0.117$
& $0.125^{+0.010}_{-0.013}$
& $0.123^{+0.009}_{-0.011}$
& $0.122^{+0.010}_{-0.013}$   
& $0.119^{+0.007}_{-0.011}$
\\ 
\hline 
  $h$   
& $0.7$
&$0.707^{+0.015}_{-0.016}$
& $0.707^{+0.012}_{-0.013}$ 
& $0.706^{+0.015}_{-0.016}$ 
& $0.702^{+0.011}_{-0.013}$
  \\ \hline
$A$   & $1$ 
& $0.941^{+0.139}_{-0.202}$ 
& $1.022^{+0.134}_{-0.184}$ 
& $1.013^{+0.153}_{-0.204}$ 
& $1.157^{+0.152}_{-0.175}$
\\ 
\hline
$n_s$  
& $0.96$
& $0.920^{+0.071}_{-0.071}$ 
& $0.941^{+0.066}_{-0.064}$
& $0.918^{+0.072}_{-0.067}$ 
& $0.950^{+0.061}_{-0.052}$
\\
\hline\hline
$\Omega_m$ 
& $0.286$ 
& $0.297^{+0.017}_{-0.019}$ 
& $0.292^{+0.014}_{-0.016}$
& $0.291^{+0.015}_{-0.018}$ 
& $0.289^{+0.011}_{-0.014}$
\\ \hline
$\sigma_8$   
& $0.82$
& $0.812^{+0.062}_{-0.067}$  
& $0.845^{+0.051}_{-0.056}$
& $0.828^{+0.058}_{-0.060}$  
& $0.885^{+0.046}_{-0.045}$
\\ 
\hline
\end{tabular}
\caption{The marginalized 1d intervals for the cosmological parameters 
			estimated from the Nseries mock data at $z_{\rm eff}=0.55$. 
			The shown are the fitted parameters (first column), fiducial values used 
			in simulations (second column), 
			the results for $P_0+P_2$ (third column) and $P_0+P_2+P_4$ (fourth column) both at $k_{\rm max}=0.20\hMpc$, along with the same combinations at $k_{\rm max}=0.25\hMpc$ (fifth and sixth columns).}
\label{tab:Nseries_all}
\end{table*}

\subsubsection{Test on PT Challenge simulations}

\begin{figure*}[ht!]
\begin{center}
\includegraphics[width=1\textwidth]{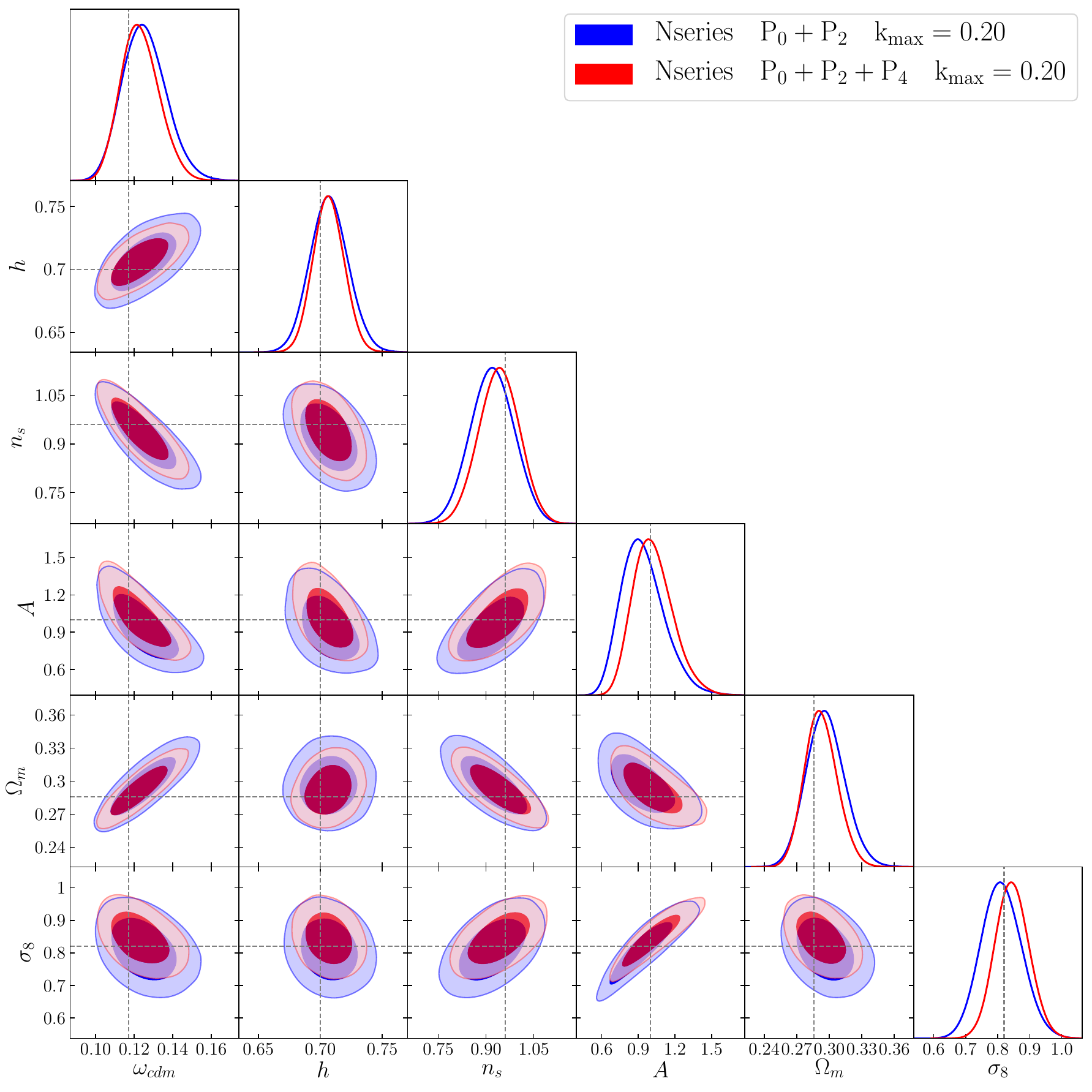}
\end{center}
\caption{
Posterior distributions of the cosmological parameters of the $\Lambda$CDM model fitted to the Nseries mock data for $\kmax=0.2\hMpc$.
\label{fig:Nseries_kmax020} } 
\end{figure*}

\begin{figure*}[ht]
\begin{center}
\includegraphics[width=1\textwidth]{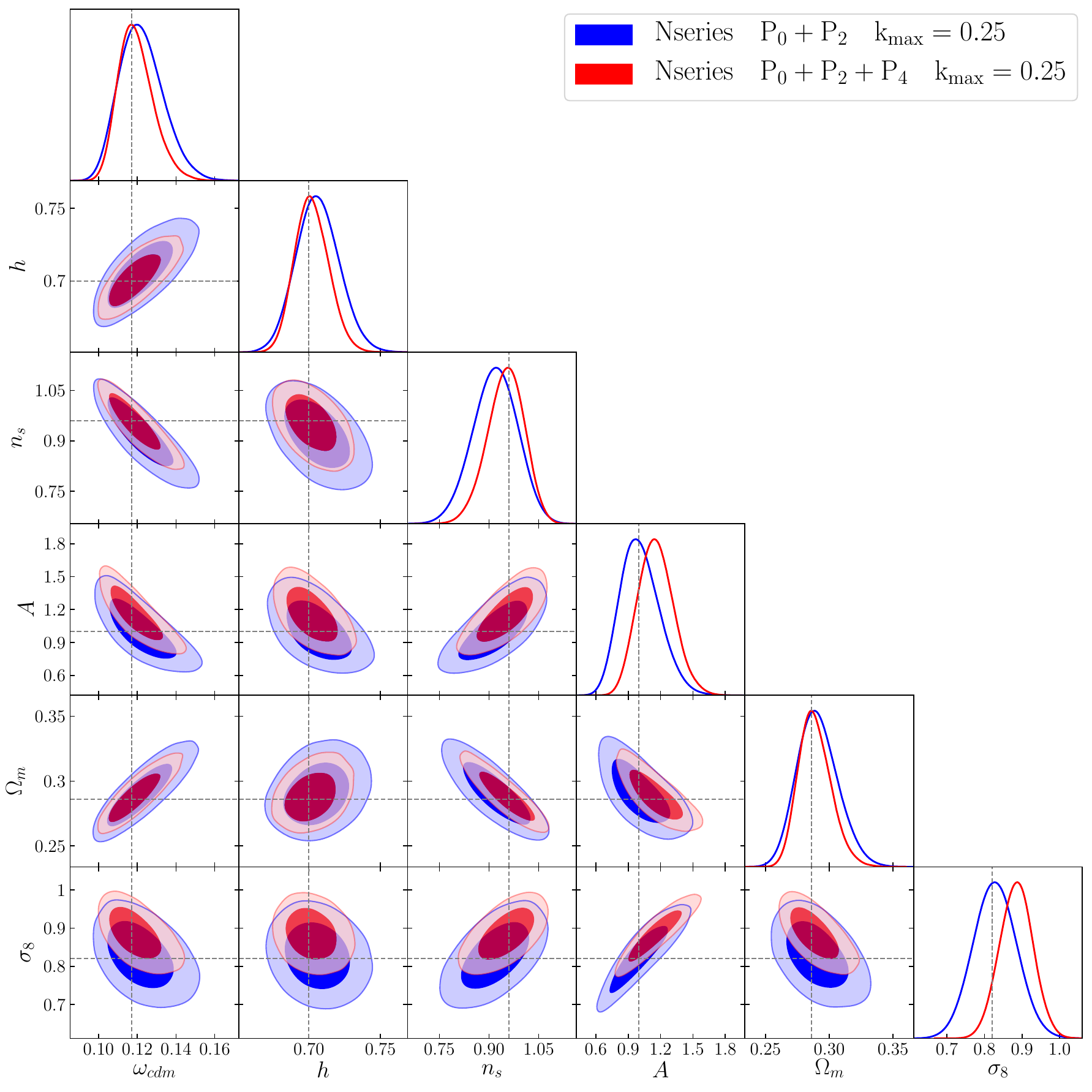}
\end{center}
\caption{
Posterior distributions of the cosmological parameters of the $\Lambda$CDM model fitted to the Nseries mock data for $\kmax=0.25\hMpc$.
\label{fig:Nseries_kmax025} } 
\end{figure*}

\begin{figure*}[ht!]
\begin{center}
\includegraphics[width=1\textwidth]{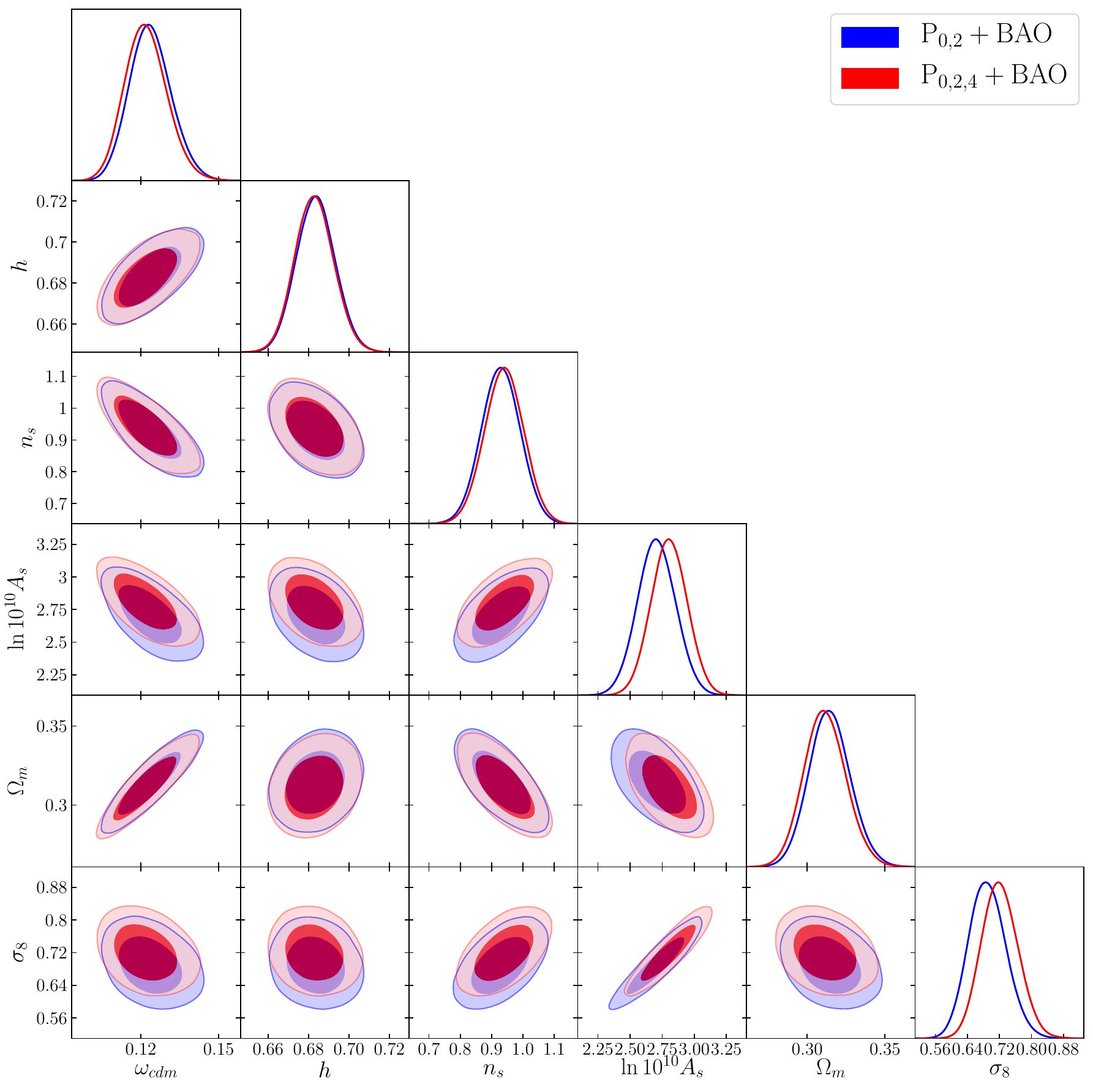}
\end{center}
\caption{
Posterior distributions of the cosmological parameters of the $\Lambda$CDM model inferred without and with the hexadecapole.
\label{fig:lcdm_hexa} } 
\end{figure*}
\begin{figure*}[ht]
\begin{center}
\includegraphics[width=1\textwidth]{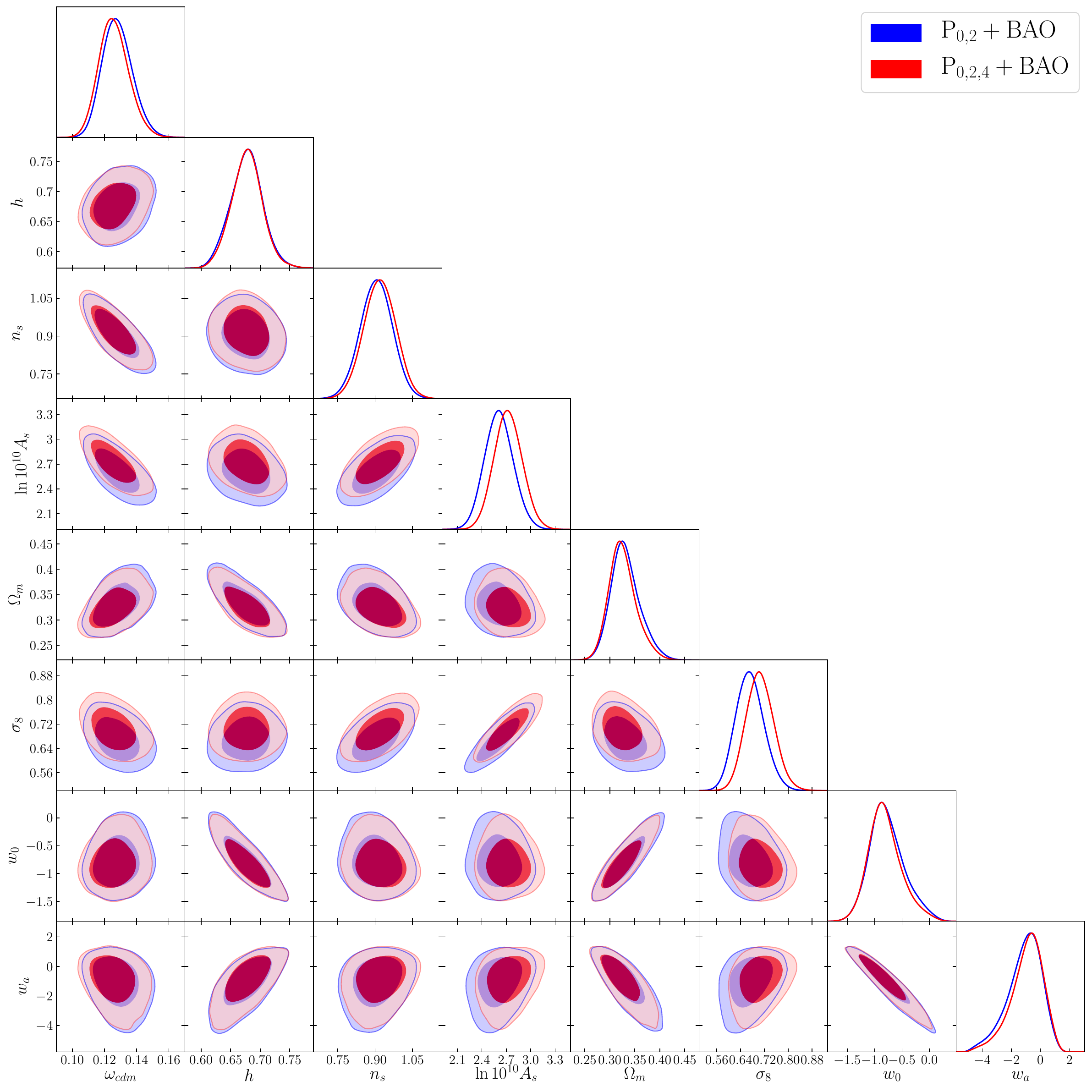}
\end{center}
\caption{
Posterior distributions of the cosmological parameters of the $w_0w_a$CDM model inferred without and with the hexadecapole.
\label{fig:w0wa_hexa} } 
\end{figure*}

The PT challenge simulation suite was designed 
for testing perturbation theory modeling at the sub-percent level. 
These N-body simulations 
reproduce the clustering of the BOSS-like galaxies 
from the DR12 sample, in a gigantic 
cumulative volume of 
$\sim 560~(\text{Gpc}/h)^3$. 

We will fit the data vector 
including the monopole, quadrupole and hexadecapole,
$\{P_0,P_2,P_4\}$ taken from a snapshot at $z=0.61$, which corresponds to the high-z NGC sample, which is the most constraining BOSS data chunk.
We use the mean data vector corresponding to the total simulation volume, but analyze it using a covariance which corresponds to the actual BOSS survey. 
In particular, we assume a Gaussian covariance for the 
power spectrum multipoles, with the following volumes 
and shot noise,
\be 
\begin{split}
&\text{BOSS-like}:~ \bar{n}^{-1}=5\cdot 10^3~[\text{Mpc}/h]^3\,,~ V=6~(\text{Gpc}/h)^3\,,\\
&\text{10x BOSS-like}:~\bar{n}^{-1}=5\cdot 10^3~[\text{Mpc}/h]^3\,,~ V=60~(\text{Gpc}/h)^3\,.
\end{split}
\ee 
The first choice corresponds to the cumulative volume of the BOSS survey. The second covariance corresponds to future surveys like DESI~\cite{Aghamousa:2016zmz} and this case provides a more stringent test of our theory model, which will also be important in order to quantify the
impact of the Bayesian parameter volume effects.
We choose the data cut $k_{\rm max}=0.2~h/$Mpc.

Our fitting model is characterized by the cosmological parameters of the base $\Lambda$CDM model,
\be 
\{\omega_{cdm},n_s,h,A_s\}\,,
\ee 
and we fix the baryon density $\omega_b$
to the fiducial value in order to simulate the BBN prior.
We will also use the following
convenient amplitude parameter 
\be 
\label{eq:A}
A\equiv \frac{A_s}{A_{s,~\text{fid}}}\,,\quad 
A_{s,~\text{fid}}=2.2109\cdot 10^{-9}
\ee 
During our MCMC analysis, we compute our theoretical templates with  \texttt{CLASS-PT}~\cite{Chudaykin:2020aoj}.
It should be mentioned that our
current 
one-loop calculation is based on the Einstein-de-Sitter approximation, 
which has been shown to be accurate
both in $\Lambda$CDM
and its extensions~\cite{Donath:2020abv}.

The results of our analysis are shown in Fig.~\ref{fig:PTC} and in Table~\ref{tab:PTC}.
As the perturbation theory challenge is still ongoing, we display the parameters
normalized to the fiducial values in order to keep the 
true cosmology blinded. 

Let us first look at the BOSS-like case. We see that our pipeline reproduces the true value of cosmological 
parameters to percent and sub-percent precision. Moreover, we find the error bars that are very similar to the actual error bars from the analysis of the full BOSS
data.
The shifts in the relevant cosmological parameters $\{\omega_{cdm},n_s,h,\ln(10^{10}A_s),\sigma_8,\Omega_m\}$ are 
\[
\{3.7,-1.5,0.6,-1.1,-0.16,1.8\}~\%,
\]
or 
\[ 
\{0.37,-0.2,0.33,-0.21,-0.027,0.27\}~\sigma
\]
if formulated in terms of the standard deviations.

Part of these shifts is produced by marginalization effects. Even though these effects are present in the data as well, it is instructive to perform an analysis 
with a smaller covariance in order to get an idea 
on their size.
For the 10xBOSS case, the shifts
are 
\[ 
\{1.5,-0.39,0.19,0.69,1.9,0.86\}\%
\] or
\[
\{0.15,-0.05,0.1,0.13,0.32,0.13\}\sigma\]
if formulated in terms of the error bars of the actual BOSS analysis. We conclude that for the survey of the BOSS volume the parameter volume effects represent the dominant part of the apparent shift of cosmological parameters from the true values. 
The marginalization effects
affect the posteriors of all cosmological parameters at the level of $0.3\sigma$.
The true theory-systematic 
shifts due to higher-order 
non-linear corrections are
very marginal, the largest one is in $\sigma_8$ and it has $0.3\sigma$ significance.

\subsubsection{Test on Nseries simulations}
\label{Nseries}

As a second test, we validate our pipeline on 
the Nseries cut-sky mock catalogs, 
which closely reproduce the actual BOSS CMASS sample (largely overlapping with the NGC high-z footprint used in our main analysis), including the appropriate survey geometry 
and selection functions.
These mocks are based on full N-body simulations and hence accurately reproduce the dynamics of gravitation clustering.
The details of the mocks are given 
in Ref.~\cite{Hand:2017pqn}.
These mocks were generated for the following fiducial cosmology: $\Omega_m= 0.286$, $\Omega_b=0.047$~($\Omega_b h^2$=0.023), $\ln(10^{10}A_s)=3.065$, $n_s$ = 0.96, $\sigma_8$=0.82, and $h$ = 0.7. The resulting spectra for Nseries have been obtained assuming
a fiducial matter abundance $\Omega_m=0.31$ when converting redshifts and angles 
into comoving distances. 
The same fiducial 
matter abundance was used 
in the actual BOSS data 
and in our theoretical templates, which 
include the Alcock-Paczynski effect. 
The effective redshift is $z_{\rm eff}=0.55$.

We use the data vector averaged over 84 cut-sky 
realizations 
to suppress the statistical fluctuations. The data is analyzed using the covariance matrix from the 
Patchy mocks, which corresponds to the BOSS CMASS NGC sample. We decided to not use Nseries mock data to build the covariance for the following two reasons. First, the Patchy mocks reproduce the BOSS CMASS NGC sample that guarantees the same parameter volume effects as the analysis of the real data. Second, Nseries sample does not include veto effect which was imprinted into the Patchy mocks and real data. Since our goal is to validate the theoretical framework on the real data, we extract the covariance from the NGC Patchy mock and not
from different realizations 
of the Nseries mocks.

All in all, the Nseries simulations reproduce general characteristics of the BOSS CMASS NGC sample, which has effective tomographic 
volume 
$2.8$~(Gpc/$h$)$^3$. The inverse number density for these mocks is 
\be 
\bar n^{-1}=5.3\cdot 10^3~[\text{Mpc}/h]^3~.
\ee 
Since the mocks have a non-trivial 
mask, we 
convolve the theoretical spectra with the survey
window function as prescribed by Ref.~\cite{Beutler:2016arn}.


The results of our analysis are shown in Fig.~\ref{fig:Nseries_kmax020} (for $k_{\rm max}=0.2~h/$Mpc),
and Fig.~\ref{fig:Nseries_kmax025}
(for $k_{\rm max}=0.25~h/$Mpc). 1d marginalized limits are given in Table~\ref{tab:Nseries_all}. 

Let us first focus on the data cut $k_{\rm max}=0.2~\hMpc$. From Fig.~\ref{fig:Nseries_kmax020} we see that our pipeline gives unbiased results
with or without the hexadecapole moment at this data cut. 
The shifts between the means and the true values of cosmological parameters
are $\lesssim 0.5\sigma$, consistent with the 
marginalization effects found earlier in the analysis of the PT challenge data. 

Now let us consider a more aggressive
data cut $k_{\rm max}=0.25~h$/Mpc.
In this case, the addition of the hexadecapole shrinks
the parameter error bars
such that the results 
at $k_{\rm max}=0.25~h$/Mpc (which was the baseline data cut in Ref.~\cite{Ivanov:2019hqk}) become biased. 
Indeed, looking at Fig.~\ref{fig:Nseries_kmax025} we see that the true cosmology is 
beyond the $95\%$ confidence interval in the 2d
space $\sigma_8-\Omega_m$,
which is not the case 
for the $P_{0,2}$ analysis.
This is the main reason why we chose a more conservative
data cut $k_{\rm max}=0.2~h$/Mpc in the baseline analysis 
of this paper.
Importantly, the error bars in the case $P_{0,2,4}$
at $k_{\rm max}=0.2~\hMpc$  
are smaller than the error bars in the case 
$P_{0,2}$ at
$k_{\rm max}=0.25~\hMpc$,
which suggests that it is 
more beneficial to include the 
hexadecapole at
$k_{\rm max}=0.2~\hMpc$ than
pushing to $k_{\rm max}=0.25~\hMpc$
with the monopole and quadrupole only.

All in all, the results of this section imply that $k_{\rm max}=0.2~\hMpc$ is a reasonable 
data cut, for which the total systematic error, including the
modeling uncertainties and
marginalization effects, 
is smaller than the statistical
error.

\subsection{Cosmological information from the hexadecapole}

\begin{table*}[t!]
  \begin{tabular}{|c||c|c||c|c|} \hline
   Param.
   & $\Lambda$CDM, $P_{0,2}$
   & $\Lambda$CDM, $P_{0,2,4}$  
   & $w_0w_a$CDM, $P_{0,2}$
   & $w_0w_a$CDM, $P_{0,2,4}$
 \\ [0.2cm]
\hline
$\omega_{cdm}$   
& $0.1237_{-0.0087}^{+0.0077}$
& $0.1221_{-0.0089}^{+0.0074}$
& $0.1283_{-0.01}^{+0.0082}$   
& $0.126_{-0.01}^{+0.0085}$ 
\\ 
\hline 
  $h$   
&$0.6836_{-0.0099}^{+0.0096}$
& $0.683_{-0.0099}^{+0.0095}$
& $0.6767_{-0.028}^{+0.027}$ 
& $0.6775_{-0.026}^{+0.026}$
  \\ \hline
$\ln(10^{10}A_s)$  
& $2.71^{+0.15}_{-0.15}$ 
& $2.80^{+0.14}_{-0.14}$ 
& $2.61^{+0.18}_{-0.18}$ 
& $2.72^{+0.17}_{-0.17}$
\\ 
\hline
$n_s$  
& $0.9305_{-0.064}^{+0.062}$ 
& $0.9401_{-0.062}^{+0.063}$
& $0.904_{-0.061}^{+0.067}$
& $0.9191_{-0.065}^{+0.066}$
\\
\hline
$w_0$  
& $-$ 
& $-$
& $-0.7618_{-0.36}^{+0.26}$
& $-0.805_{-0.34}^{+0.25}$
\\
\hline
$w_a$  
& $-$ 
& $-$
& $-1.141_{-0.84}^{+1.4}$
& $-0.9451_{-0.83}^{+1.3}$
\\
\hline\hline
$\Omega_{\rm de}$ 
& $-$ 
& $-$
& $0.668^{+0.033}_{-0.024}$ 
& $0.673^{+0.031}_{-0.022}$
\\ \hline
$\Omega_m$ 
& $0.3145_{-0.014}^{+0.013}$
& $0.3116_{-0.014}^{+0.013}$
& $0.3322_{-0.033}^{+0.024}$ 
& $0.3262_{-0.031}^{+0.023}$
\\ \hline
$\sigma_8$   
& $0.691^{+0.043}_{-0.050}$  
& $0.720^{+0.042}_{-0.047}$
& $0.673^{+0.043}_{-0.050}$  
& $0.705^{+0.044}_{-0.049}$
\\ \hline
$r_d$ [Mpc]  & $145_{-2.4}^{+2.4}$
& $145_{-2.4}^{+2.4}$
& $145_{-2.4}^{+2.4}$ 
& $145_{-2.4}^{+2.4}$\\
\hline
\end{tabular}
\caption{The marginalized 1d intervals for the cosmological parameters 
			estimated from the BBN+FS+BAO likelihood. 
			We show the results for $\Lambda$CDM (second and third columns) and 
			for the $w_0w_a$CDM model 
			(fourth and fifth columns).
			In either case we display the results obtained with and without
			the hexadecapole moment.
			}
\label{tab:hexa}
\end{table*}

It is instructive 
to quantify how much the power
spectrum hexadecapole improves
the parameter constraints 
compared to the monopole and quadrupole combination.  
To that end we analyze the BOSS FS data using two different data vectors, $P_{0}+P_{0}$ and 
$P_{0}+P_{2}+P_4$, both at $k_{\rm max}=0.2~\hMpc$, and
including
the BAO data along with the BBN prior 
on $\omega_b$.

Let us first focus on the base $\Lambda$CDM model. The results of our analysis are displayed in Fig.~\ref{fig:lcdm_hexa} and in Table~\ref{tab:hexa}.
We see that the hexadecapole 
does not noticeably 
improve the parameter constraints. 
The only result of the inclusion the hexadecapole moment is a marginal upward shift of $\sigma_8$. The same tendency has been found in the results with the Nseries mocks which reliable describe the survey geometry and selection functions, see Sec. \ref{Nseries}. 
This effect can be attributed to 
the reduction of the parameter volume effect, which also brings the amplitude closer to the Planck prediction. 
All in all, we find the inclusion of the hexadecapole reduces the total volume in cosmological parameter space by $2\%$ compared to the monopole and quadrupole analysis in the base
$\Lambda$CDM model.

Now let us focus on the $w_0w_a\Lambda$CDM model. This model has a larger number of free parameters and hence we expect the gain from 
the hexadecapole to be more significant here. However, looking at Fig.~\ref{fig:w0wa_hexa} and in Table~\ref{tab:hexa}, 
we see that the improvement is quite marginal here.
This result should be contrasted with the claims of Ref.~\cite{Beutler:2016arn} that the hexadecapole improves
the constraints on the distance parameters inferred through
the so-called alpha-analysis. 
Having repeated this analysis we have indeed reproduced the same $\sim 30\%$ improvement with our theoretical model and scale cuts. 
However, we see that this gain
does not propagate into the actual physical parameters even 
in the extended $w_0w_a\Lambda$CDM model. 
It remains to be seen if the improvement 
from the hexadecapole reported in Ref.~\cite{Beutler:2016arn}
is merely an artifact of the scaling alpha-analysis, 
which does not correspond to any 
physical model.  

Even though the effect of the hexadecapole on the 1d parameter constraints
is quite marginal, it should be mentioned that it decreases
the total volume of the 2d posterior $w_0-w_a$ by $19\%$ as compared 
to the $P_{0,2}$ combination.
The addition of the hexadecapole
seems to be more beneficial 
if we consider the total posterior
volume of all sampled cosmological
parameters, which
reduces by $26\%$.
Thus, the 1d marginalized limits might 
not fully reflect the information
content of the hexadecapole.

\section{Combined analyses with Planck}
\label{app:planck}

\begin{figure*}[ht]
\begin{center}
\includegraphics[width=1\textwidth]{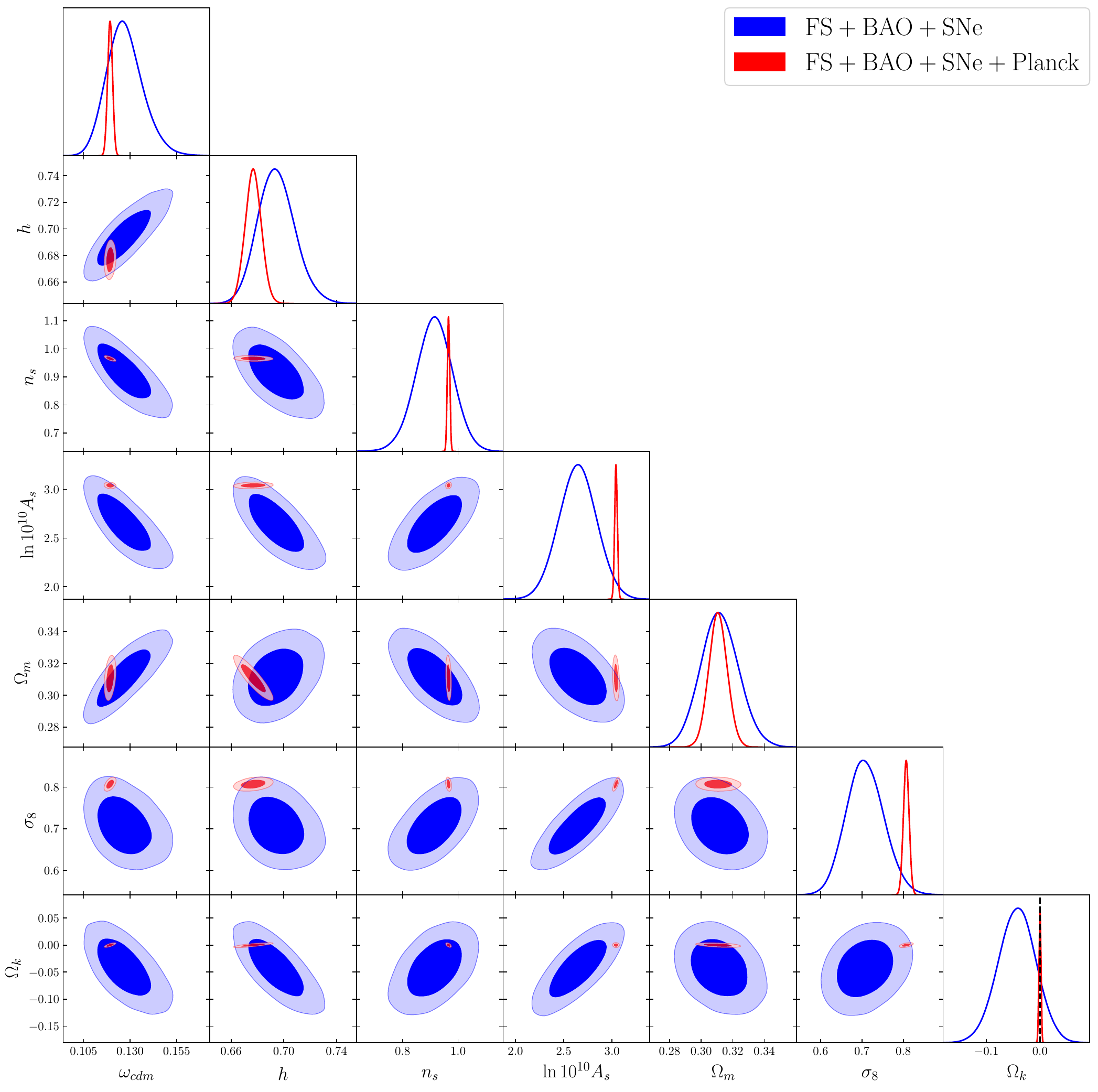}
\end{center}
\caption{
Posterior distributions of the cosmological parameters of the o$\Lambda$CDM model.
\label{fig:Ok_planck} } 
\end{figure*}

\begin{figure*}[ht]
\begin{center}
\includegraphics[width=1\textwidth]{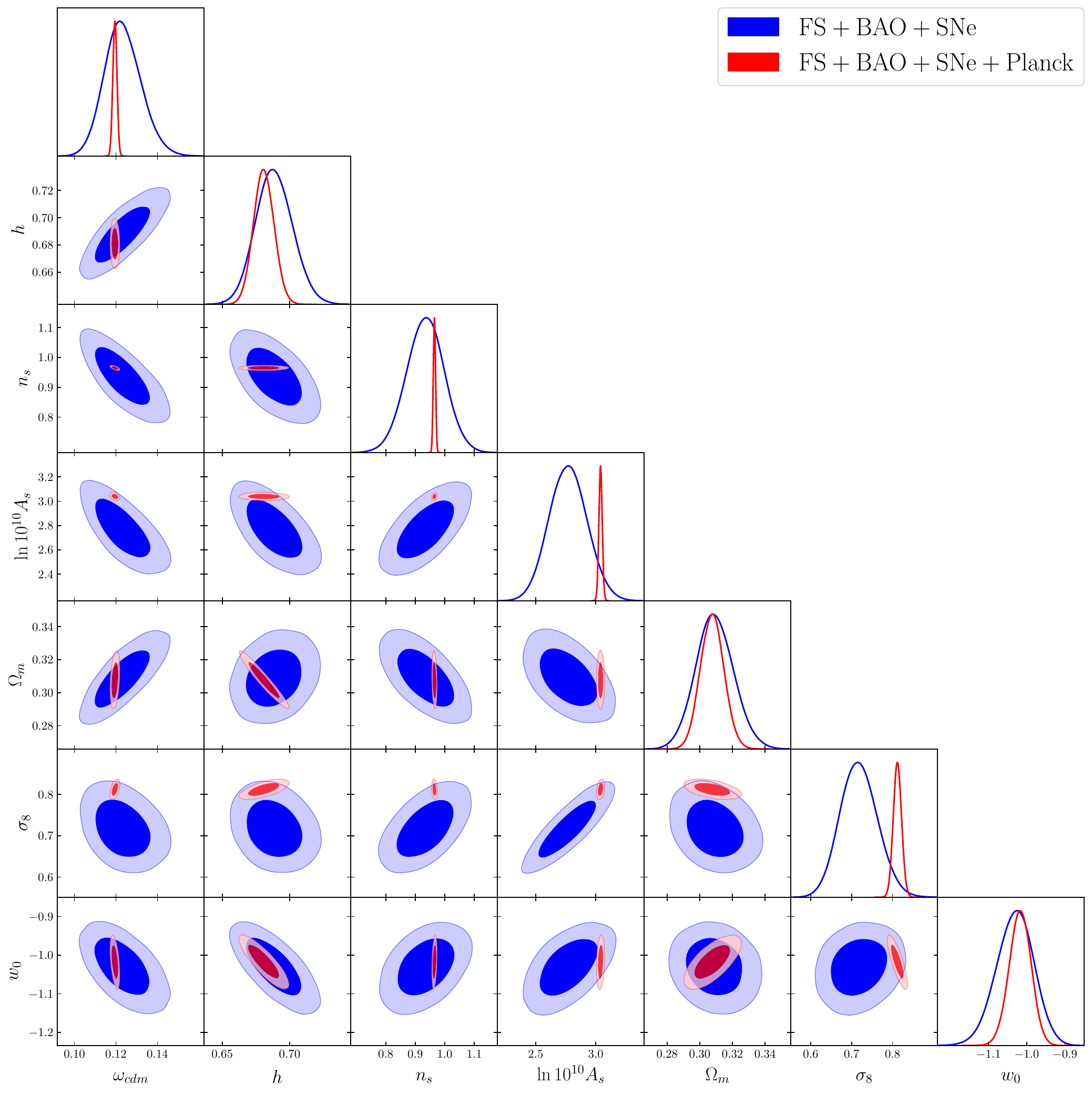}
\end{center}
\caption{
Posterior distributions of the cosmological parameters of the $w_0$CDM model.
\label{fig:w0_planck} } 
\end{figure*}

\begin{figure*}[ht]
\begin{center}
\includegraphics[width=1\textwidth]{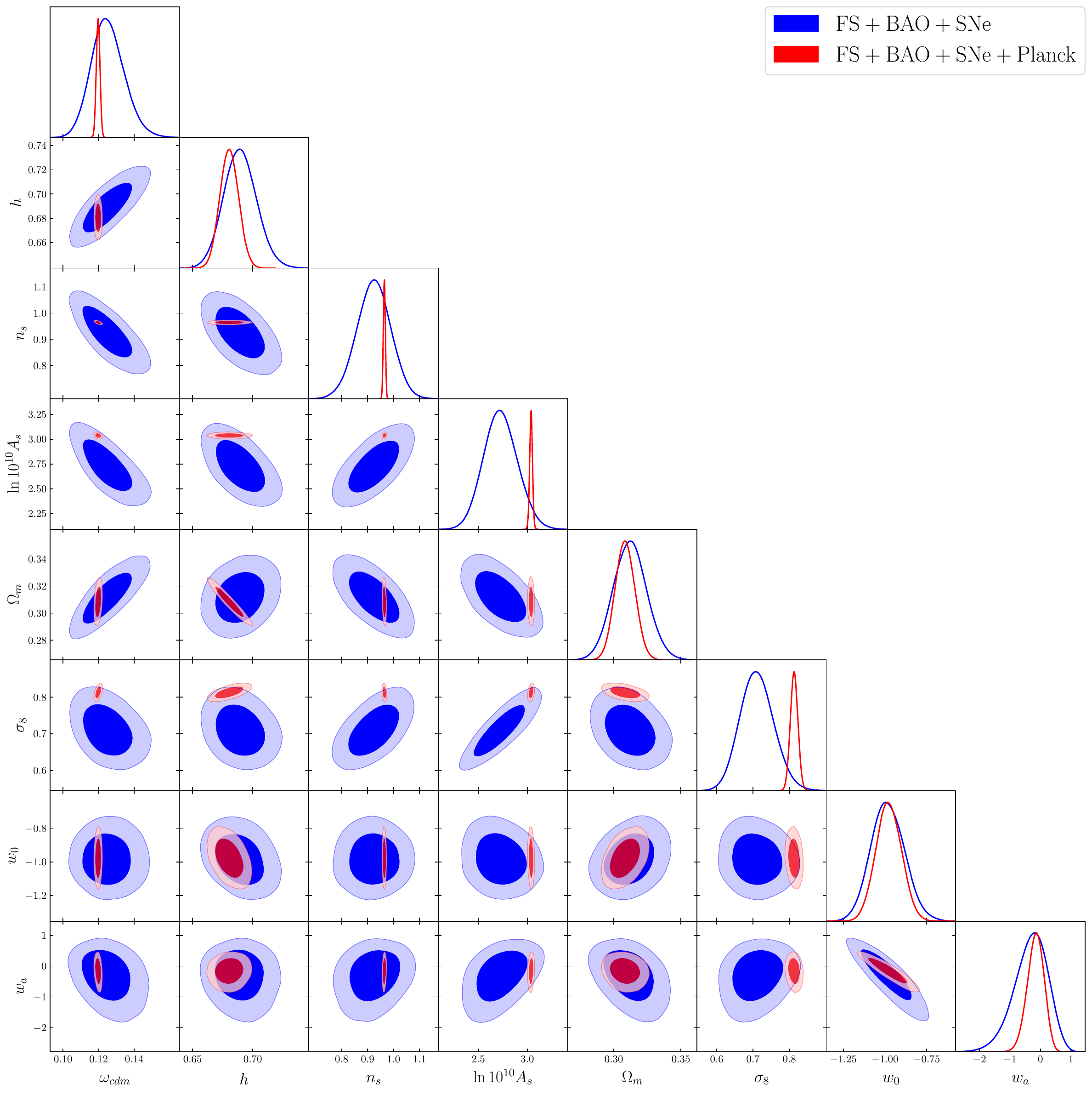}
\end{center}
\caption{
Posterior distributions of the cosmological parameters of the $w_0w_a$CDM model.
\label{fig:w0wa_planck} } 
\end{figure*}

In this Appendix we 
present the results of the joint analysis of the FS, BAO, SNe, and Planck CMB likelihoods.
We use the Planck baseline likelihood $\rm TT,TE,EE+lowE+lensing$ in the notation of Ref. \cite{Aghanim:2018eyx} that comprises temperature, polarisation and lensing potential power spectra measurements. 
The 1d marginalized constraints for 
o$\Lambda$CDM, $w_0$CDM and $w_0w_a$CDM models
are
reported in the 3rd, 5th and 7th columns of Tab. \ref{table2}, along with our baseline results (without Planck)
that are displayed 
in the 2nd, 4th and 6th columns, respectively. The corresponding triangle plots for the parameters are shown in Fig.~\ref{fig:Ok_planck}, \ref{fig:w0_planck} and \ref{fig:w0wa_planck}.

\begin{table*}[t!]
  \begin{tabular}{|c||c|c||c|c||c|c|} \hline
   \diagbox{ {\small Param.}}{\small Dataset}  
   & {\small FBS, o$\Lambda$CDM }  
   & {\small FBSP, o$\Lambda$CDM }  
   &  FBS, $w_0$CDM   
   &  FBSP, $w_0$CDM   
   & FBS, $w_0w_a$CDM 
   & FBSP, $w_0w_a$CDM 
      \\ [0.2cm]
\hline
$\omega_{cdm}$   
& $0.1273_{-0.011}^{+0.0086}$
& $0.1192_{-1.3\cdot10^{-3}}^{+1.3\cdot10^{-3}}$
& $0.1233_{-0.0096}^{+0.0083}$
& $0.1195_{-1.0\cdot10^{-3}}^{+1.0\cdot10^{-3}}$
& $0.1252_{-0.01}^{+0.0085}$
& $0.1197_{-1.1\cdot10^{-3}}^{+1.1\cdot10^{-3}}$
\\ 
\hline
$10^2\omega_{b}$   
& $2.267_{-0.039}^{+0.039}$
& $2.241_{-0.015}^{+0.015}$
& $2.266_{-0.038}^{+0.039}$
& $2.239_{-0.014}^{+0.013}$
& $2.266_{-0.038}^{+0.038}$
& $2.237_{-0.014}^{+0.014}$
\\ 
\hline 
$h$   
& $0.6945_{-0.015}^{+0.013}$
& $0.6768_{-6.2\cdot10^{-3}}^{+6.3\cdot10^{-3}}$
& $0.6885_{-0.014}^{+0.013}$
& $0.6808_{-7.9\cdot10^{-3}}^{+7.6\cdot10^{-3}}$
& $0.6896_{-0.014}^{+0.013}$
& $0.6805_{-8.0\cdot10^{-3}}^{+7.9\cdot10^{-3}}$
\\ \hline
$\ln(10^{10}A_s)$   
& $2.65_{-0.20}^{+0.20}$ 
& $3.041_{-0.014}^{+0.015}$ 
& $2.77_{-0.16}^{+0.16}$
& $3.039_{-0.014}^{+0.014}$ 
& $2.72_{-0.17}^{+0.17}$
& $3.037_{-0.015}^{+0.015}$ 
\\ 
\hline
$n_s$  
& $0.9136_{-0.065}^{+0.068}$
& $0.9656_{-4.5\cdot10^{-3}}^{+4.4\cdot10^{-3}}$
& $0.9346_{-0.064}^{+0.065}$
& $0.9649_{-4.0\cdot10^{-3}}^{+3.9\cdot10^{-3}}$
& $0.9236_{-0.065}^{+0.066}$
& $0.9644_{-4.0\cdot10^{-3}}^{+4.0\cdot10^{-3}}$
\\
\hline
$\tau$  
& $-$
& $0.0536_{-7.3\cdot10^{-3}}^{+7.2\cdot10^{-3}}$
& $-$
& $0.0528_{-7.3\cdot10^{-3}}^{+7.2\cdot10^{-3}}$
& $-$
& $0.0518_{-7.5\cdot10^{-3}}^{+7.4\cdot10^{-3}}$
\\
\hline
$\Omega_k$   
& $-0.043_{-0.036}^{+0.036}$
& $0.0000_{-1.8\cdot10^{-3}}^{+2.0\cdot10^{-3}}$
& $-$
& $-$ 
& $-$
& $-$
\\   \hline 
$w_0$   
& $-$ 
& $-$
& $-1.031_{-0.048}^{+0.052}$
& $-1.018_{-0.028}^{+0.030}$
& $-0.983_{-0.11}^{+0.099}$ 
& $-0.978_{-0.078}^{+0.079}$ 
\\   \hline
$w_a$   
& $-$ 
& $-$
& $-$
& $-$ 
& $-0.326_{-0.479}^{+0.629}$
& $-0.163_{-0.259}^{+0.301}$
\\   \hline
\hline
$\Omega_\Lambda$   
& $0.731_{-0.033}^{+0.033}$
& $0.6893_{-5.3\cdot10^{-3}}^{+5.6\cdot10^{-3}}$
& $-$
& $-$ 
& $-$ 
& $-$ 
\\ \hline
$\Omega_m$   
&  $0.3119_{-0.013}^{+0.012}$
&  $0.3107_{-6.0\cdot10^{-3}}^{+5.9\cdot10^{-3}}$
& $0.3091_{-0.012}^{+0.011}$
& $0.3077_{-7.6\cdot10^{-3}}^{+7.2\cdot10^{-3}}$
& $0.3121_{-0.013}^{+0.012}$
&  $0.3084_{-7.7\cdot10^{-3}}^{+7.5\cdot10^{-3}}$
\\ \hline
$\Omega_{\rm de}$   
& $-$
& $-$
& $0.691_{-0.011}^{+0.011}$
& $0.6924_{-7.1\cdot10^{-3}}^{+7.2\cdot10^{-3}}$
& $0.688_{-0.012}^{+0.012}$
& $0.6916_{-7.4\cdot10^{-3}}^{+7.3\cdot10^{-3}}$
\\ \hline
$\sigma_8$   
&  $0.708_{-0.048}^{+0.043}$
&  $0.8069_{-6.9\cdot10^{-3}}^{+6.8\cdot10^{-3}}$
& $0.718_{-0.048}^{+0.043}$ 
& $0.812_{-0.011}^{+0.010}$
& $0.711_{-0.049}^{+0.044}$
&  $0.813_{-0.011}^{+0.011}$
\\ 
\hline
$r_d$ [Mpc]  
& $146_{-2.4}^{+2.4}$
& $144.70_{-0.29}^{+0.29}$
& $146_{-2.4}^{+2.4}$ 
& $144.64_{-0.24}^{+0.23}$
& $146_{-2.4}^{+2.4}$
& $144.60_{-0.24}^{+0.24}$
\\ 
\hline
\end{tabular}
\caption{Mean values and 68\% CL minimum credible
intervals for the parameters of the various extended models for two data sets differing by the presence of the supernovae data. FBS denotes the combination FS+BAO+SNe, FBSP denotes the combination FS+BAO+SNe+Planck. The BBN prior on $\omega_b$ is assumed only in FBS analyses.
The top group represents the parameters that were directly varied in the MCMC chains. The bottom group
are the derived parameters.}
\label{table2}
\end{table*}

Let us begin with the o$\Lambda$CDM model. 
There Planck 
improves the constraints
on all cosmological
parameters, especially on
the spatial curvature, which we find to be
$\Omega_k=0.0000_{-1.8\cdot10^{-3}}^{+2.0\cdot10^{-3}}$.
It is also worth noting that the mass fluctuation amplitude $\sigma_8$ inferred from FS+BAO+SNe+Planck is  higher (at $2\sigma$ C.L.) than our baseline FS+BAO+SNe estimate,
which reflects 
the so-called 
$\sigma_8$-tension \cite{DiValentino:2020vvd}.

Now let us focus on the $w_0$CDM model. In this scenario, including the Planck data yields a significant improvement for the shape and amplitude parameters, $\omega_{cdm},n_s,A_s$ and $\sigma_8$, whereas the gain for the background parameters, $w_0$ and $H_0$, is quite modest. In particular, our FS+BAO+SNe+Planck analysis yields $w_0=-1.018_{-0.028}^{+0.030}$ which improves the FS+BAO+SNe estimate only by $40\%$. 

Finally, we examine the 
cosmological parameters 
of the $w_0w_a$CDM model. The situation 
here is similar to the $w_0$CDM 
scenario. 
The Planck data 
strengthens the bounds 
on the shape and 
amplitude parameters,
but does not dramatically
affect the background
expansion parameters.
In particular, 
the FS+BAO+SNe+Planck analysis yields $w_0=-0.978_{-0.078}^{+0.079}$ that represents a merely $20\%$ improvement over the FS+BAO+SNe estimate. 
The improvement for $w_a$
is more sizeable, about a factor of two,
$w_a=-0.163_{-0.259}^{+0.301}$.

\bibliography{short.bib}

\end{document}